\documentclass[journal]{IEEEtran}
\usepackage{amsmath,amsfonts}
\usepackage{algorithmic}
\usepackage{algorithm}
\usepackage{array}
\usepackage[caption=false,font=normalsize,labelfont=sf,textfont=sf]{subfig}
\usepackage{textcomp}
\usepackage{stfloats}
\usepackage{url}
\usepackage{verbatim}
\usepackage{graphicx}
\usepackage{cite}
\usepackage{xtab}
\usepackage{bbm}
\usepackage{booktabs}
\usepackage{xcolor}
\begin{document}

\title{MLFCIL: A Multi-Level Forgetting Mitigation Framework for Federated Class-Incremental Learning in LEO Satellites}

\author{
    Heng Zhang,
    Xiaohong Deng, 
    Sijing Duan,~\IEEEmembership{Member,~IEEE,}
    Wu Ouyang,  \\
    KM Mahfujul,~\IEEEmembership{Member,~IEEE,}
    Yiqin Deng,~\IEEEmembership{Member,~IEEE,}
    Zhigang Chen,~\IEEEmembership{Member,~IEEE}
    \thanks{This work was supported in part by the National Natural Science Foundation of China Project under Grant No. 61762046, Grant 62166019, Grant 62301300, and 62402279, in part by the Postdoctoral Fellowship Program and China Postdoctoral Science Foundation under Grant No. BX20250375, in part by the Jiangxi Provincial Natural Science Foundation Joint Fund-Key Project under Grant 20253BAC280098, in part by the China Postdoctoral Science Foundation under Grant No. 2025M771503, and in part by the Jiangxi Provincial Department of Education Science and Technology Research Project under Grant GJJ2503509. (Corresponding authors: Xiaohong Deng and Sijing Duan.)}
    \thanks{Heng Zhang, Xiaohong Deng, and Wu Ouyang are with the School of Information Engineering, Gannan University of Science and Technology, Ganzhou 341000, China (e-mail: zhangheng@gnust.edu.cn; 9320070286@gnust.edu.cn; ouyangwu@gnust.edu.cn).}
    \thanks{Sijing Duan is with the Department of Computer Science and Technology, Tsinghua University, Beijing 100084, China (e-mail: duansj@tsinghua.edu.cn).}
    \thanks{KM Mahfujul is with the Department of Electrical and Computer Engineering, Queen University, Kingston, ON K7L 3N6, Canada (e-mail: m.kadir@queensu.ca).}
    \thanks{Yiqin Deng is with School of Data Science, Lingnan University, Tuen Mun, Hong Kong, China (email: yiqindeng@ln.edu.hk).}
    \thanks{Zhigang Chen is with the School of Computer Science and Engineering, Central South University, Changsha 410083, China (e-mail: czg@csu.edu.cn).}
}

\markboth{Journal of \LaTeX\ Class Files,~Vol.~14, No.~8, August~2021}%
{Shell \MakeLowercase{\textit{et al.}}: A Sample Article Using IEEEtran.cls for IEEE Journals}

\IEEEpubid{0000--0000/00\$00.00~\copyright~2021 IEEE}

\maketitle

\begin{abstract}
    Low-Earth-orbit (LEO) satellite constellations are increasingly performing on-board computing. However, the continuous emergence of new classes under strict memory and communication constraints poses major challenges for collaborative training. Federated class-incremental learning (FCIL) enables distributed incremental learning without sharing raw data, but faces three LEO-specific challenges: non-independent and identically distributed data heterogeneity caused by orbital dynamics, amplified catastrophic forgetting during aggregation, and the need to balance stability and plasticity under limited resources. To tackle these challenges, we propose MLFCIL, a multi-level forgetting mitigation framework that decomposes catastrophic forgetting into three sources and addresses them at different levels: class-reweighted loss to reduce local bias, knowledge distillation with feature replay and prototype-guided drift compensation to preserve cross-task knowledge, and class-aware aggregation to mitigate forgetting during federation. In addition, we design a dual-granularity coordination strategy that combines round-level adaptive loss balancing with step-level gradient projection to further enhance the stability–plasticity trade-off. Experiments on the NWPU-RESISC45 dataset show that MLFCIL significantly outperforms baselines in both accuracy and forgetting mitigation, while introducing minimal resource overhead.
\end{abstract}

\begin{IEEEkeywords}
Federated class-incremental learning, catastrophic forgetting, LEO satellite, remote sensing.
\end{IEEEkeywords}

\section{Introduction}
\label{sec:introduction}

\IEEEPARstart{L}{ow}-Earth-orbit (LEO) satellite constellations are increasingly equipped with onboard computing capabilities, enabling GPU-accelerated satellites to perform sensing, inference, and learning directly in space. This reduces reliance on ground stations and mitigates limitations such as intermittent contact windows and limited downlink bandwidth~\cite{ref1,ref2,ref3,ref4}. As a result, satellites can support applications such as land-cover classification, object detection, and change detection using onboard imagery ~\cite{ref5,ref6,ref7,ref8}. However, as constellations grow and tasks become more diverse, the computing capacity of individual satellites may become insufficient~\cite{ref9}. This motivates collaborative approaches where multiple satellites coordinate their learning processes through Federated Learning (FL), which enables distributed intelligence across the constellation~\cite{ref10,ref11}. In particular, FL enables satellites to collaboratively train a shared model while keeping observation data onboard and respecting the bandwidth limitations of inter-satellite links (ISLs)~\cite{ref12,ref13,ref14,ref15,ref16,ref17}.

\begin{figure}[!t]
    \centering
    \includegraphics[width=\columnwidth]{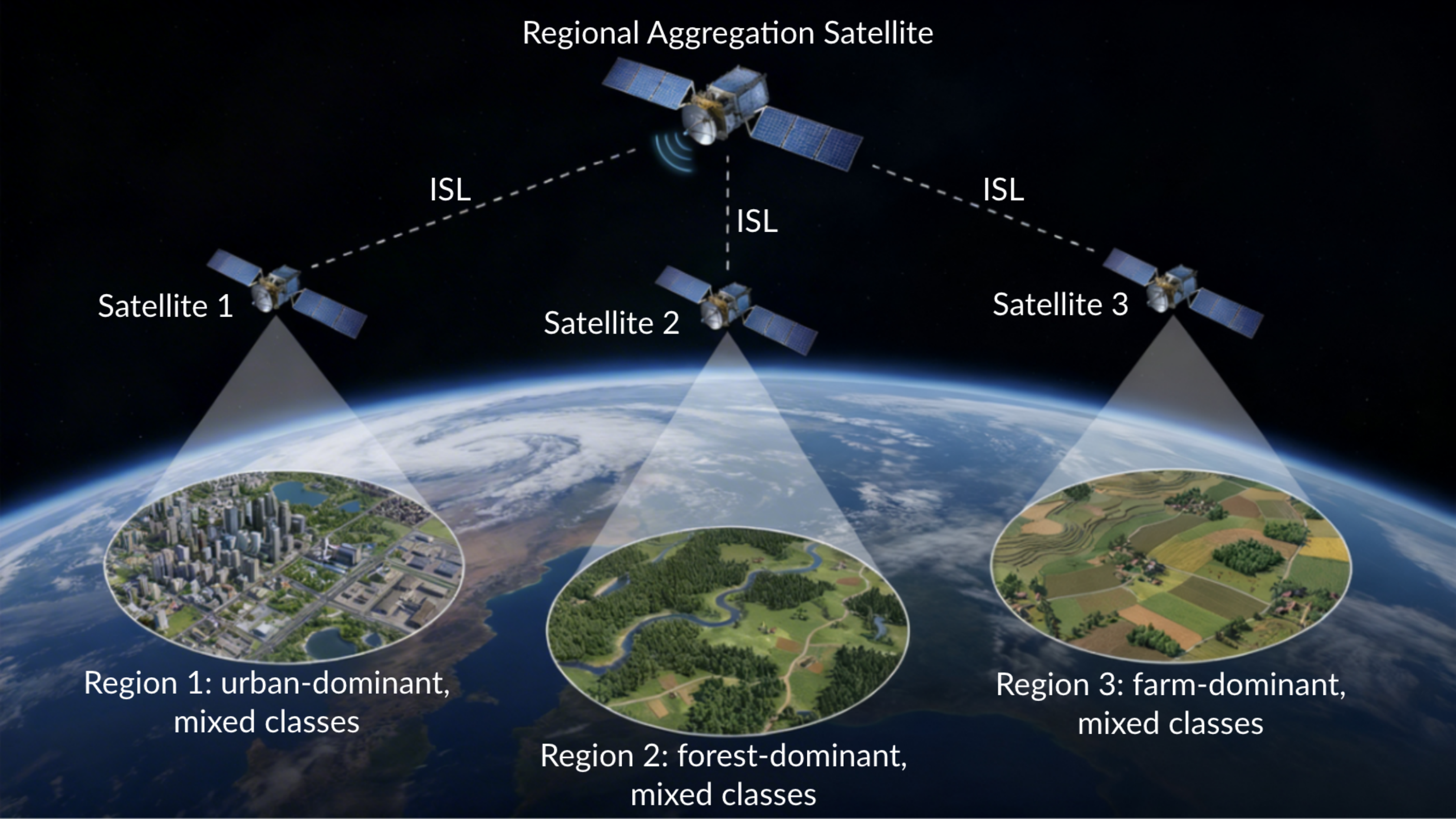}
    \caption{FCIL scenario in LEO: edge satellites perform on-board class-incremental training and synchronize with a regional aggregation satellite via inter-satellite links (ISL), without reliance on ground stations.}
    \label{fig:scenario}
\end{figure}

However, Earth observation in LEO is inherently dynamic. As satellites traverse different orbital regions, they continuously encounter new land-cover types and object categories that may not be present in the initial training set~\cite{ref18}. This leads to evolving and heterogeneous label spaces across satellites (e.g., non-independent identically distributed data), making the fixed-label assumption of conventional FL unsuitable~\cite{ref19,ref20,ref21,ref22}. To address this limitation, Federated Class-Incremental Learning (FCIL) has recently emerged as a framework that enables satellites to collaboratively learn new classes over time while retaining previously acquired knowledge~\cite{ref23,ref24,ref25}.  As illustrated in Fig.~\ref{fig:scenario}, multiple LEO edge satellites perform onboard incremental learning over region-specific data and synchronize model updates with a regional aggregation satellite (a compute- and storage-enhanced LEO satellite) via inter-satellite links, entirely without relying on ground stations. Despite its potential, FCIL faces several key challenges in LEO environments. First, orbit-dependent observations result in heterogeneous label spaces and asynchronous task arrivals across satellites. Second, aggregating model updates from clients with evolving class sets can exacerbate catastrophic forgetting and introduce aggregation bias. Third, maintaining the stability–plasticity balance while adapting to new classes is significantly constrained by limited onboard memory, computation, and ISL bandwidth

To address these challenges, this paper proposes \textbf{MLFCIL}, a multi-level forgetting mitigation framework for FCIL in LEO satellites. MLFCIL addresses forgetting at three complementary levels. First, it enforces class-balanced learning during incremental training to ensure newly arriving classes are effectively learned. Second, it incorporates knowledge preservation mechanisms that maintain consistency with previously learned tasks and mitigate representation drift for existing features. Third, it introduces a class-aware federated aggregation strategy specifically tailored to heterogeneous class distributions across satellites, preventing rare classes from being overshadowed during global model updates. In addition, MLFCIL employs a dual-granularity coordination mechanism that balance adaptation to new classes with the preservation of prior knowledge under the resource constraints of LEO satellites. Experiments on the NWPU-RESISC45 benchmark under non-independent identically distributed (Non-IID) settings show that MLFCIL outperforms federated baselines with minimal resource overhead.

The main contributions are as follows:
\begin{itemize}
    \item We analyze catastrophic forgetting in LEO federated class-incremental learning and decompose it into three sources: local training bias, where each satellite’s model is skewed towards its own non-IID data. Inter-task knowledge loss, where learning new tasks overwrites representations of old ones. And aggregation-induced forgetting amplification, where server-side model aggregation further magnifies forgetting under heterogeneous label spaces. Together, these three sources provide a structured foundation for multi-level mitigation.
    
    \item We propose MLFCIL, a multi-level forgetting mitigation framework that addresses the above challenges through class-reweighted loss, knowledge distillation with feature-level memory replay, prototype-guided drift compensation, class-aware federated aggregation, and a dual-granularity coordination strategy. 
    
    \item We evaluate MLFCIL on the NWPU-RESISC45 benchmark under Non-IID satellite scenarios, demonstrating superior performance over federated baselines in both accuracy and forgetting mitigation. 
\end{itemize}

The remainder of this paper is organized as follows: Section~\ref{sec:related_work} reviews related work. Section~\ref{sec:scenario_problem} describes the system scenario and problem formulation. Section~\ref{sec:methodology_design} presents the MLFCIL methodology. Section~\ref{sec:experiments} provides an experimental evaluation. Section~\ref{sec:conclusion} concludes the paper.

\section{Related Work}
\label{sec:related_work}

This section reviews relevant work organized around the three aspects: Non-IID data heterogeneity in FL, catastrophic forgetting in class-incremental learning (CIL), and FCIL in LEO satellites.

\subsection{Non-IID Data Heterogeneity in FL}

FL enables collaborative model training across distributed clients without centralizing raw data~\cite{ref12,ref13,ref14,ref26}. A central challenge is \emph{statistical heterogeneity}: when local data distributions differ across clients (Non-IID), the global model can converge slowly or reach a suboptimal solution~\cite{ref13}. The literature categorizes Non-IID heterogeneity into label distribution skew, feature distribution skew, and quantity skew~\cite{ref12,ref13}. Existing remedies include robust aggregation strategies that adjust server-side model averaging~\cite{ref12}, variance-reduction and proximal-regularization techniques that correct for client drift~\cite{ref13}, and personalized FL methods that allow locally adapted model components. In LEO satellite networks, heterogeneity is particularly severe because satellites on different orbital planes observe geographically distinct regions with fundamentally different class distributions~\cite{ref12}. Lin et al.~\cite{ref15} proposed FedSN, employing sub-network training and pseudo-synchronous aggregation to handle model staleness caused by intermittent connectivity. Zhai et al.~\cite{ref16} introduced FedLEO, a decentralized FL framework with task offloading to address computational heterogeneity. Razmi et al.~\cite{ref27} demonstrated on-board FL for dense LEO constellations that leverages inter-satellite links for in-network aggregation, achieving gains in convergence speed and communication efficiency.

However, these methods all assume a \emph{fixed} label space and do not address the temporal evolution of observation data. In evolving scenarios, heterogeneous aggregation further \emph{amplifies} catastrophic forgetting across incremental tasks, motivating the class-aware aggregation mechanism in MLFCIL.

\subsection{Catastrophic Forgetting in CIL}

CIL requires a model to sequentially learn new categories while retaining knowledge of previously learned ones, all within a unified classifier that does not rely on task identifiers at inference time~\cite{ref20,ref28,ref29}. Existing strategies can be broadly grouped into regularization-based methods that constrain updates to parameters important for old tasks, replay-based methods that retain or generate exemplars of past data, and architecture-based methods that allocate dedicated capacity to each task~\cite{ref28,ref30}. More recently, prototype-based approaches that maintain compact class-mean representations in the embedding space have gained traction, particularly in resource-constrained settings where storing raw exemplars is impractical~\cite{ref20,ref28,ref31}. A central concern is the \emph{stability-plasticity} tradeoff: excessive stability preserves old knowledge at the cost of poor new-class adaptation, while excessive plasticity accelerates forgetting~\cite{ref19,ref21}. Dohare et al.~\cite{ref19} demonstrated that continual training can even cause a progressive \emph{loss of plasticity}, underscoring the importance of mechanisms that explicitly balance both directions. In remote sensing, Zhao et al.~\cite{ref5} addressed forgetting and class imbalance via STCL-DRNet, Wei et al.~\cite{ref6} developed a class-incremental framework with explicit bias correction for scene classification, and Zhou et al.~\cite{ref18} combined contrastive learning with angular penalty loss for HSI classification.

These methods operate in single-device settings and do not address the distributed, heterogeneous data landscape of multi-satellite constellations. Extending CIL to such environments requires handling Non-IID heterogeneity and catastrophic forgetting simultaneously, which is precisely the setting addressed by federated class-incremental learning.

\subsection{FCIL in LEO satellites}

FCIL extends CIL to the federated setting, where multiple clients collaboratively learn incrementally arriving classes without sharing raw data~\cite{ref22,ref24,ref32,ref33}. Beyond classical CIL forgetting, FCIL introduces additional challenges: \emph{spatial forgetting}, where heterogeneous local updates across clients dilute globally shared knowledge during aggregation, and \emph{temporal forgetting}, where successive task transitions erode previously consolidated representations~\cite{ref23,ref34}. In the general FCIL setting, Dong et al.~\cite{ref23} addressed heterogeneous clients through local-global anti-forgetting mechanisms that combine category-balanced gradient compensation with prototype-based augmentation. Yang et al.~\cite{ref35} proposed TARGET, an exemplar-free distillation approach that transfers knowledge from a frozen global teacher to local models without storing old-class exemplars. For \emph{LEO satellites} specifically, Niu et al.~\cite{ref25} proposed FCIL-MSN for multisatellite networks with on-satellite incremental updates. Overall, FCIL is a natural fit for LEO satellites that continuously encounter new observation categories and require a single global model that performs well on all classes seen so far.

Despite this progress, FCIL research tailored to the LEO satellite scenario remains limited, and existing FCIL methods generally address local and global forgetting in isolation, without jointly coordinating class-aware aggregation, adaptive loss balancing, and gradient-level stability-plasticity management within a unified framework. The proposed MLFCIL framework is designed to bridge this gap.

\section{System Scenario and Problem Formulation}
\label{sec:scenario_problem}

This section describes the LEO satellite FCIL scenario, formulates the studied problem, and defines the optimization objective. Table~\ref{tab:notation} summarizes the key mathematical notations.

\subsection{System Scenario}
\label{sec:system_scenario}

\begin{figure}[!t]
    \centering
    \includegraphics[width=\columnwidth]{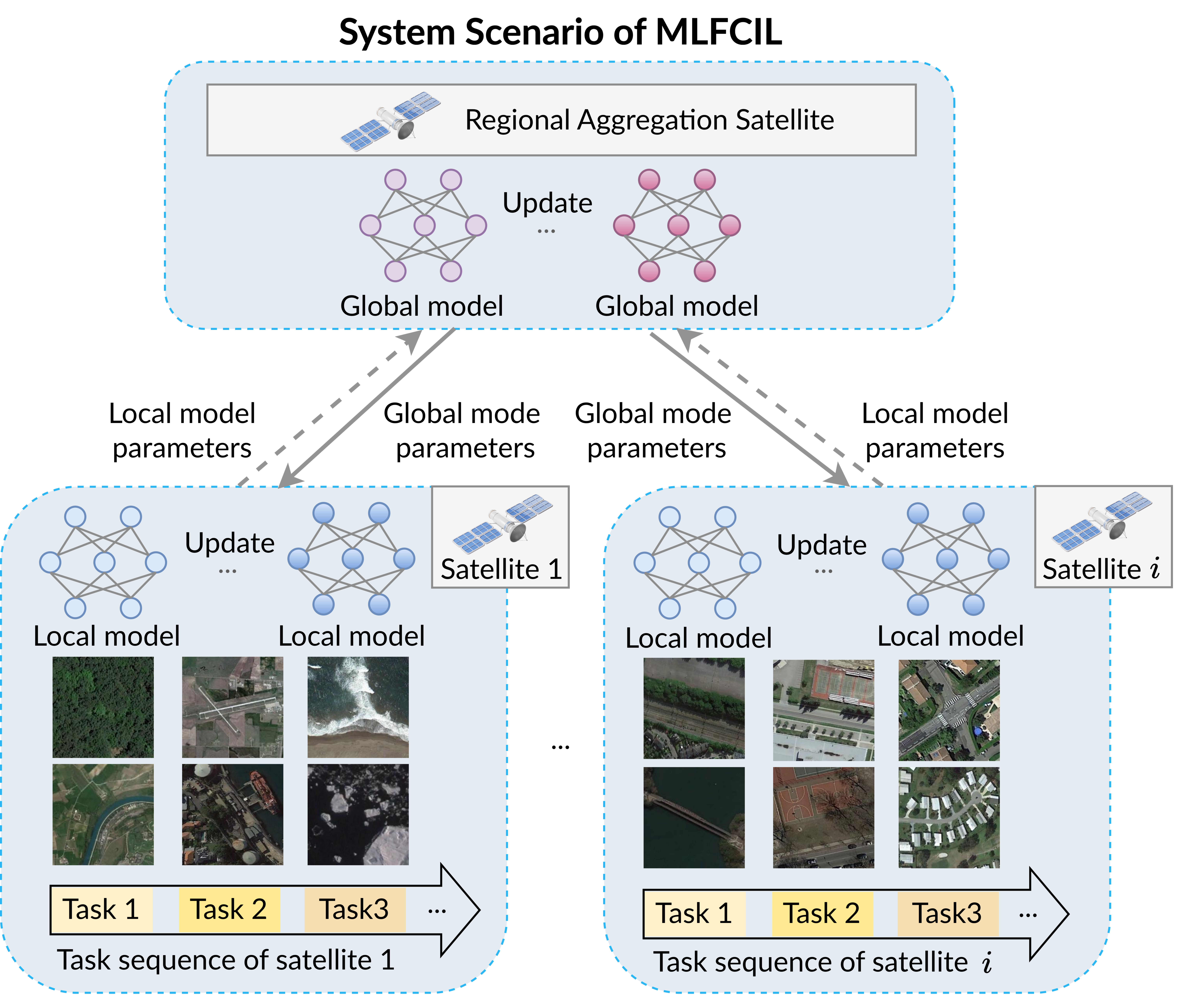}
    \caption{System scenario of MLFCIL.}
    \label{fig:system_scenario}
\end{figure}

As illustrated in Fig.~\ref{fig:system_scenario}, the considered scenario consists of a regional aggregation satellite and $N$ satellite edge nodes (Satellite 1, \ldots, Satellite $N$) connected via inter-satellite links. Each edge node continuously acquires region-specific remote sensing imagery and processes a sequence of class-incremental tasks (Task 1, Task 2, Task 3, \ldots), where new observation categories emerge as orbital coverage shifts over time. The regional aggregation satellite serves as a parameter server that coordinates collaborative model training across the constellation. In each communication round, the aggregation satellite broadcasts the current global model parameters to all edge nodes. Each edge node initializes its local model from the received parameters, trains on its current task data, and returns the updated local model parameters to the aggregation satellite. The aggregation satellite then combines the collected updates to produce the next global model. Lightweight metadata, including class prototypes and per-class sample statistics, are also exchanged alongside model parameters. Raw data never leaves any edge node, preserving data privacy and respecting the limited bandwidth of inter-satellite links.


\begin{table}[!t]
    \renewcommand{\arraystretch}{1.15}
    \caption{Key Mathematical Notations}
    \label{tab:notation}
    \centering
    \footnotesize
    \begin{tabular}{@{}l@{\hspace{8pt}}p{\dimexpr\columnwidth-3.3cm\relax}@{}}
    \toprule
    \textbf{Notation} & \textbf{Description} \\
    \midrule
    $N$, $T$, $C$ & Number of clients, tasks, and classes. \\
    $\mathcal{N}$, $\mathcal{Y}_t$, $\mathcal{Y}_{1:t}$ & Client index set; class set of task $t$; cumulative class set. \\
    $\mathcal{D}_i^{(t)}$ & Local dataset at client $i$ for task $t$. \\
    $n_i^{(c)}$, $N_c$ & Class-$c$ samples at client $i$; total $N_c = \sum_i n_i^{(c)}$. \\
    \midrule
    $\theta{=}\{\theta_f, \theta_c\}$ & Model: feature extractor $\theta_f$ and classifier $\theta_c$. \\
    $\theta_i$, $\theta_G$, $\theta_{\text{old}}$ & Local model at client $i$; global model; teacher model from previous task. \\
    $\mathbf{z} \in \mathbb{R}^d$ & Feature embedding ($d$-dimensional). \\
    $\mathbf{p}_c^{(i)}$, $\mathbf{p}_c^{(G)}$, $\hat{\mathbf{p}}_c^{(t)}$ & Local/global prototype; snapshot at task $t$. \\
    $\boldsymbol{\delta}_c$ & Drift vector for replay compensation. \\
    \midrule
    $\mathcal{M}_i$, $M_{\max}$ & Memory replay buffer; maximum buffer size. \\
    $\tilde{\mathbf{z}}$ & Drift-compensated feature embedding. \\
    $\alpha_i^{(c)}$, $w_c$ & Per-class aggregation weight; class weight for loss. \\
    \midrule
    $\mathcal{L}_{\text{cls}}$, $\mathcal{L}_{\text{distill}}$, $\mathcal{L}_{\text{replay}}$ & Classification, distillation, and replay losses. \\
    $\lambda_{\text{distill}}^{(r)}$, $\lambda_{\text{replay}}^{(r)}$ & Adaptive loss weights at round $r$. \\
    $\mathcal{F}_i^{(r)}$ & Forgetting score at client $i$ in round $r$. \\
    $\mathbf{g}_{\text{plas}}$, $\mathbf{g}_{\text{stab}}$, $\mathbf{g}_{\text{plas}}^{\prime}$ & Plasticity / stability gradients on $\theta_f$; projected plasticity. \\
    \midrule
    $E_G$, $E_L$, $\eta$ & Global rounds; local epochs; learning rate. \\
    $\tau$ & Distillation temperature, typically $\tau{=}2$. \\
    $A_t$, PD & Cumulative accuracy; performance degradation. \\
    \bottomrule
    \end{tabular}
\end{table}

\subsection{Problem Formulation}
\label{sec:problem_formulation}

We formalize the above scenario as a federated class-incremental learning problem. Let $\mathcal{N} = \{1, 2, \ldots, N\}$ index the satellite edge nodes and $\mathcal{Y} = \{0, 1, \ldots, C-1\}$ denote the global label space with $C$ classes. Unlike conventional FL where all classes are available from the start, classes arrive sequentially in disjoint subsets, forming a task sequence.

\textbf{Task Sequence.} The learning process is divided into $T$ tasks. For each task $t$, a subset of classes $\mathcal{Y}_t \subset \mathcal{Y}$ is introduced, such that:
\begin{equation}
\mathcal{Y}_t \cap \mathcal{Y}_{t'} = \emptyset, \quad \forall t \neq t', \quad \text{and} \quad \bigcup_{t=1}^{T} \mathcal{Y}_t = \mathcal{Y}.
\label{eq:class_disjoint}
\end{equation}
The cumulative class set at task $t$ is $\mathcal{Y}_{1:t} = \bigcup_{s=1}^{t} \mathcal{Y}_{s}$. At task $t$, each client $i$ has access to a local dataset $\mathcal{D}_i^{(t)} = \{(x_j^{(i,t)}, y_j^{(i,t)})\}_{j=1}^{n_i^{(t)}}$, where $x_j^{(i,t)} \in \mathbb{R}^{3 \times H \times W}$ is an optical image of height $H$ and width $W$, $y_j^{(i,t)} \in \mathcal{Y}_t$ is the class label, and $n_i^{(t)} = |\mathcal{D}_i^{(t)}|$ is the local dataset size. Let $\mathcal{D}_{\text{train}}$ and $\mathcal{D}_{\text{test}}$ denote the global training and test sets, respectively. For each task $t$, the training set is $\mathcal{D}_{\text{train}}^{(t)} = \{(x, y) \in \mathcal{D}_{\text{train}} \mid y \in \mathcal{Y}_t\}$ and the cumulative test set is $\mathcal{D}_{\text{test}}^{(1:t)} = \{(x, y) \in \mathcal{D}_{\text{test}} \mid y \in \mathcal{Y}_{1:t}\}$. Each task dataset is further partitioned among $N$ clients using a Dirichlet-based Non-IID allocation (details in Section~\ref{sec:experiments}).

\textbf{Non-IID Data Distribution.} Satellite remote sensing data exhibits significant geographic and temporal heterogeneity. The local data distribution $P_i(x, y)$ at client $i$ differs from both the global distribution $P(x, y)$ and distributions at other clients:
\begin{equation}
P_i(x, y) \neq P_j(x, y), \quad \text{and} \quad P_i(x, y) \neq P(x, y).
\label{eq:non_iid}
\end{equation}
This heterogeneity arises from geographic bias (different orbital paths yield different land-cover distributions), temporal bias (seasonal and diurnal shifts), and sensor variability (differences in illumination and imaging conditions).

\subsection{Optimization Objective}
\label{sec:optimization_objective}

The global model is parameterized by $\theta = \{\theta_f, \theta_c\}$, where $\theta_f$ is the feature extractor and $\theta_c$ is the classifier. The objective is to find parameters $\theta^*$ that minimize the cumulative classification error over all tasks seen so far, while preserving performance on old classes (stability) and adapting to new classes (plasticity):
\begin{equation}
\theta^* = \arg\min_{\theta} \sum_{i=1}^{N} \sum_{s=1}^{t} \mathbb{E}_{(x,y) \sim P_i^{(s)}} \left[ \ell(f_{\theta}(x), y) \right] + \lambda \mathcal{R}(\theta),
\label{eq:objective}
\end{equation}
where $f_{\theta}(\cdot)$ denotes the model, $\ell(\cdot, \cdot)$ is the loss function, $P_i^{(s)}$ is the data distribution at client $i$ for task $s$, and $\mathcal{R}(\theta)$ is a regularization term (e.g., knowledge distillation) with weight $\lambda$. The key challenge is that past task data $\{\mathcal{D}_i^{(s)}\}_{s<t}$ is no longer accessible during task $t$, making direct minimization of the cumulative objective infeasible. The framework design in Section~\ref{sec:methodology_design} addresses this through multi-level forgetting mitigation mechanisms.

\section{Methodology Design}
\label{sec:methodology_design}

\subsection{Overview of MLFCIL}
\label{sec:overview_mlfcil}

\begin{figure}[t]
    \centering
    \includegraphics[width=\columnwidth]{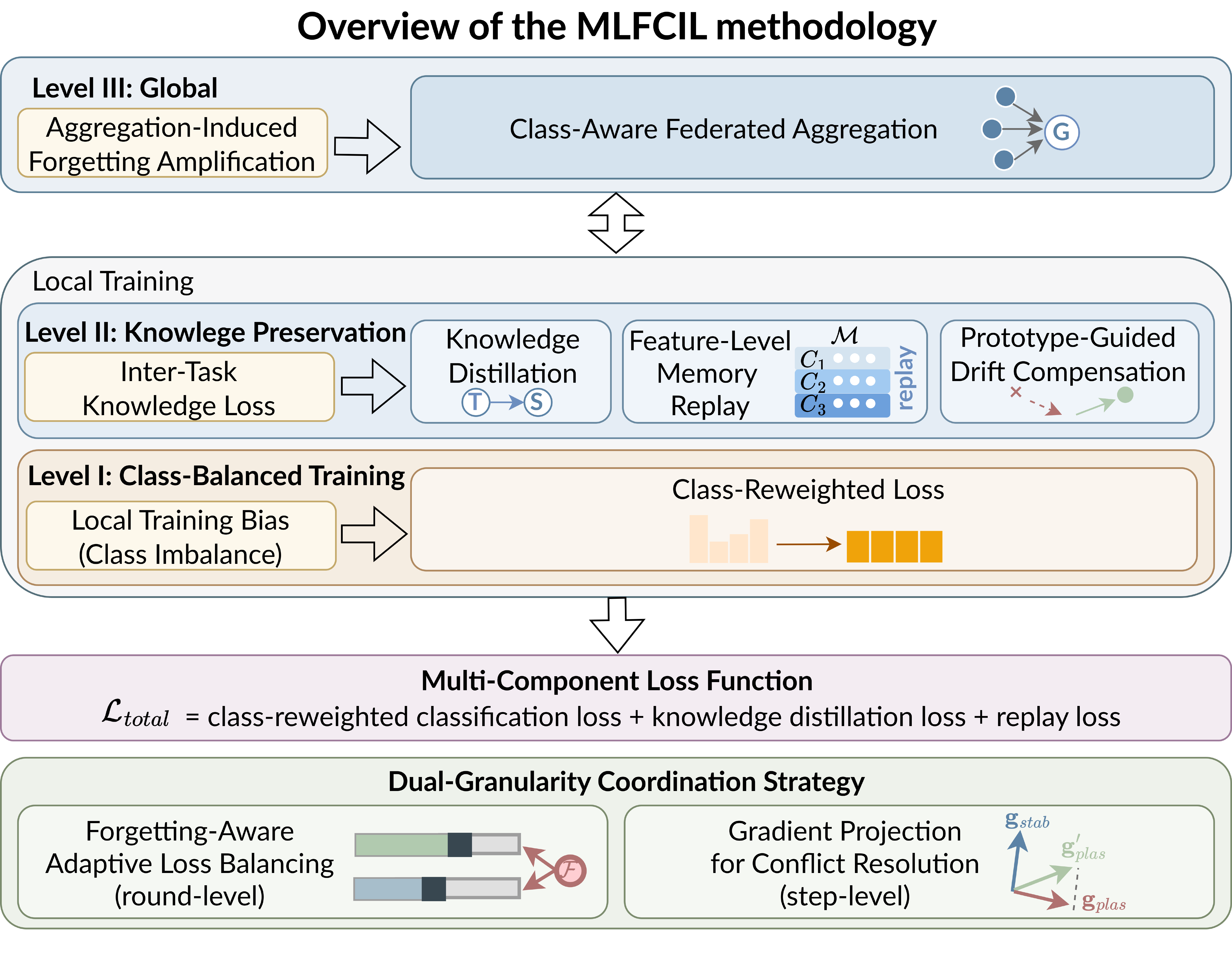}
    \caption{Overview of the MLFCIL.}
    \label{fig:methodology_overview}
\end{figure}

As illustrated in Fig.~\ref{fig:methodology_overview}, MLFCIL mitigates catastrophic forgetting through three complementary levels. The first two operate within a unified \emph{Local Training} scope on each satellite. At the \emph{class-balanced training level} (Level~I), per-class reweighting is applied based on local class statistics (Section~\ref{sec:class_balanced_training}) to address training bias caused by class imbalance. At the \emph{knowledge preservation level} (Level~II), knowledge distillation, feature-level memory replay, and prototype-guided drift compensation (Section~\ref{sec:knowledge_preservation}) are employed to maintain inter-task consistency without storing raw data. While both levels reside in local training, they address distinct objectives: Level~I ensures equitable optimization across old and new classes, whereas Level~II explicitly retains previously learned representations. These mechanisms are integrated via a multi-component loss function (Section~\ref{sec:loss_function}) that combines class-reweighted classification, knowledge distillation, and replay losses into a unified objective. At the \emph{global level} (Level~III), class-aware federated aggregation (Section~\ref{sec:federated_mechanism}) weights each satellite's contribution per class by local sample count, preventing dilution of rare-class decision boundaries. 

As indicated by the bidirectional arrow in Fig.~\ref{fig:methodology_overview}, locally trained updates propagate to global aggregation, and the aggregated model is broadcast back to guide subsequent local training. A dual-granularity coordination strategy (Section~\ref{sec:dual_granularity}) further enhances the stability-plasticity balance through round-level adaptive loss balancing and step-level gradient projection. This strategy is depicted as a separate layer, as it modulates loss weights and resolves gradient conflicts on~$\theta_f$ without introducing an additional data-flow path.

\subsection{Local Training}
\label{sec:local_training}
On each satellite, local training encompasses two complementary levels that address orthogonal aspects of catastrophic forgetting: class-balanced training (Level~I) and knowledge preservation (Level~II).

\subsubsection{Class-Balanced Training (Level~I)}
\label{sec:class_balanced_training}

The incremental learning manager tracks which classes are new ($\mathcal{Y}_{\text{new}}^{(i)}$) and which have been seen ($\mathcal{Y}_{\text{seen}}^{(i)}$) at each client $i$, and uses this to compute class-reweighted loss that emphasizes newly introduced classes. When a sample with label $y$ is encountered, if $y \notin \mathcal{Y}_{\text{seen}}^{(i)}$, then $y$ is added to both $\mathcal{Y}_{\text{new}}^{(i)}$ and $\mathcal{Y}_{\text{seen}}^{(i)}$.

\textbf{Class Weight Calculation.} To handle class imbalance and emphasize new classes, we compute per-class weights using an inverse frequency strategy with an additional boost for new classes:
\begin{equation}
w_c = \begin{cases}
\frac{\sum_{c'} n_i^{(c')}}{|\mathcal{Y}_{\text{seen}}^{(i)}| \cdot n_i^{(c)}} \cdot \beta, & \text{if } c \in \mathcal{Y}_{\text{new}}^{(i)}, \\
\frac{\sum_{c'} n_i^{(c')}}{|\mathcal{Y}_{\text{seen}}^{(i)}| \cdot n_i^{(c)}}, & \text{otherwise},
\end{cases}
\label{eq:class_weights}
\end{equation}
where $\beta > 1$ (typically $\beta = 1.5$) is a boosting factor for new classes. This weighting balances loss contributions across classes and improves plasticity without overwhelming the gradients.

\subsubsection{Knowledge Preservation (Level~II)}
\label{sec:knowledge_preservation}
The knowledge preservation level maintains inter-task consistency through three mechanisms: online prototype learning, feature-level memory replay with drift compensation, and knowledge distillation.

\textbf{Online Prototype Learning.}
Class prototypes serve as compact representations of class-specific feature distributions in the embedding space. For each class $c \in \mathcal{Y}_{1:t}$, we maintain a prototype vector $\mathbf{p}_c^{(i)} \in \mathbb{R}^d$, where $d$ is the feature dimension of the penultimate layer. During training at client $i$, when a sample $(x, y)$ with label $y = c$ is processed, we extract its feature representation:
\begin{equation}
\mathbf{z} = h_{\theta_f^{(i)}}(x) \in \mathbb{R}^d,
\label{eq:feature_extraction}
\end{equation}
where $h_{\theta_f^{(i)}}(\cdot)$ denotes the feature extractor. The class prototype is then updated via an online update rule:
\begin{equation}
\mathbf{p}_c^{(i)} \leftarrow \frac{n_i^{(c)} \cdot \mathbf{p}_c^{(i)} + \mathbf{z}}{n_i^{(c)} + 1}, \quad n_i^{(c)} \leftarrow n_i^{(c)} + 1,
\label{eq:prototype_update}
\end{equation}
where $n_i^{(c)}$ denotes the number of class-$c$ samples observed at client $i$. The resulting local prototypes $\{\mathbf{p}_c^{(i)}\}$ are uploaded to the server for federated aggregation (Section~\ref{sec:federated_mechanism}).

\textbf{Feature-Level Memory Replay Buffer.}
Naive experience replay that stores raw images would quickly exhaust on-board storage due to the large size of satellite imagery (e.g., a single $256 \times 256$ RGB image requires $\sim$200~KB). To address this, we employ \emph{feature-level replay}: each client $i$ maintains a memory buffer $\mathcal{M}_i = \{(\mathbf{z}_j, y_j)\}_{j=1}^{M_i}$, where $\mathbf{z}_j \in \mathbb{R}^d$ is a stored feature embedding and $y_j \in \mathcal{Y}_{1:t}$ the corresponding label. The total budget is $M_{\max}$ embeddings. To ensure balanced class representation, the per-class allocation is dynamically adjusted:

\begin{equation}
M_c = \max\left(5, \left\lfloor \frac{M_{\max}}{|\mathcal{Y}_{1:t}|} \right\rfloor \right),
\label{eq:samples_per_class}
\end{equation}
where $M_c$ is the maximum number of samples stored for each class, and $|\mathcal{Y}_{1:t}|$ is the number of classes seen so far. When a new sample $(x, y)$ with $y = c$ arrives:
\begin{itemize}
    \item \textbf{Step~1:} Extract feature embedding: $\mathbf{z} = h_{\theta_f^{(i)}}(x)$.
    \item \textbf{Step~2:} If the buffer for class $c$ is not full ($|\{(\mathbf{z}_j, y_j) \in \mathcal{M}_i : y_j = c\}| < M_c$), add $(\mathbf{z}, c)$ directly.
    \item \textbf{Step~3:} Otherwise, use reservoir sampling: with probability $M_c / n_i^{(c)}$, randomly replace one existing sample of class $c$ with $(\mathbf{z}, c)$.
\end{itemize}

This strategy maintains a class-balanced subset of historical samples for replay. Storing feature embeddings instead of raw images reduces memory consumption by approximately two orders of magnitude, and concrete savings are quantified in Section~\ref{sec:resource_overhead}.

\begin{figure}[t]
\centering
\includegraphics[width=0.48\textwidth]{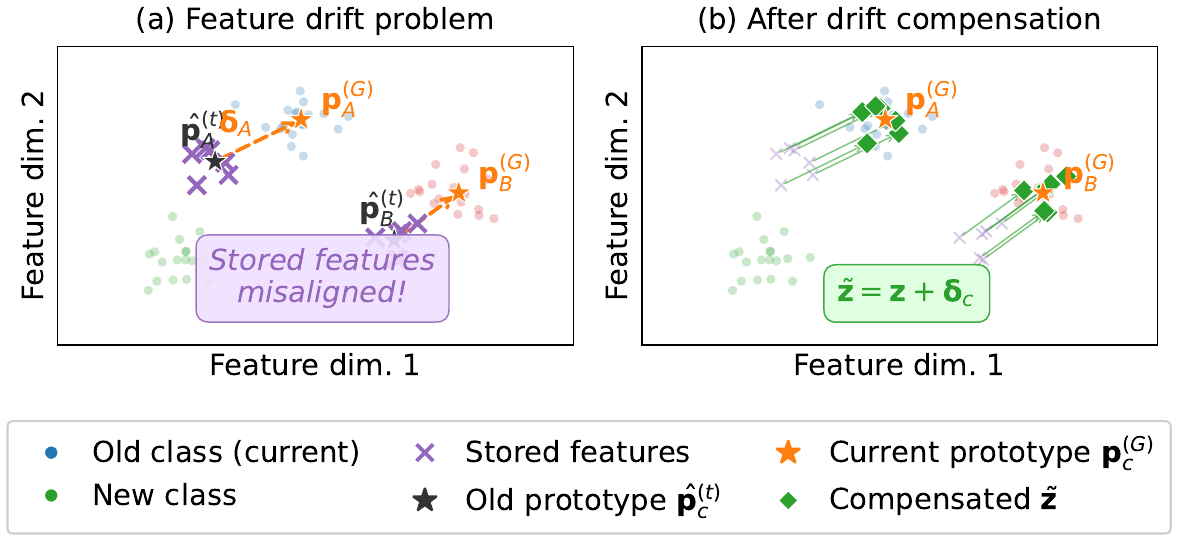}
\caption{Prototype-guided feature drift compensation. (a)~Feature drift problem. (b)~After drift compensation.}
\label{fig:drift_concept}
\end{figure}

\textbf{Prototype-Guided Feature Drift Compensation.}
A well-known limitation of feature-level replay is \emph{feature drift}: as the feature extractor $\theta_f$ is updated across tasks, previously stored embeddings become misaligned with the current feature space (Fig.~\ref{fig:drift_concept}). To address this, we propose a lightweight correction mechanism that leverages the class prototypes already maintained in the framework to estimate and compensate for this drift without additional storage or communication overhead.

At the transition from task $t$ to task $t{+}1$, each client records a snapshot of the global prototypes $\{\hat{\mathbf{p}}_c^{(t)}\}_{c \in \mathcal{Y}_{1:t}}$ (the prototypes at the time features were stored). During replay in task $t{+}1$, the current global prototypes $\{\mathbf{p}_c^{(G)}\}$ reflect the updated feature space. For a stored embedding $(\mathbf{z}_j, y_j)$ from class $c = y_j$, we apply a prototype-guided correction:
\begin{equation}
\tilde{\mathbf{z}}_j = \mathbf{z}_j + \underbrace{\left(\mathbf{p}_c^{(G)} - \hat{\mathbf{p}}_c^{(t)}\right)}_{\text{drift vector } \boldsymbol{\delta}_c},
\label{eq:feature_compensation}
\end{equation}
where $\boldsymbol{\delta}_c = \mathbf{p}_c^{(G)} - \hat{\mathbf{p}}_c^{(t)}$ estimates the class-specific drift direction in the embedding space. The corrected feature $\tilde{\mathbf{z}}_j$ is then used in place of $\mathbf{z}_j$ in the replay loss (Eq.~\eqref{eq:replay_loss}).

Under incremental fine-tuning, the feature space undergoes approximately uniform shifts within each class cluster, and translating stored features by $\boldsymbol{\delta}_c$ re-aligns them with the current space. The additional storage is $|\mathcal{Y}_{1:t}| \times d$ floating-point values (e.g., 45\,KB), and the computation is element-wise addition, both negligible for on-board deployment.

\textbf{Knowledge Distillation from Task-Wise Teachers.}
We adopt knowledge distillation to preserve previously learned decision boundaries. At each task transition $t \to t+1$, the current model is saved as the teacher $f_{\theta_{\text{old}}}$. For a given input $x$, let $\mathbf{q}_{\text{old}} = \text{softmax}(f_{\theta_{\text{old}}}(x)/\tau)$ be the teacher's soft label and $\mathbf{q}_{\text{new}} = \text{softmax}(f_{\theta_i}(x) / \tau)$ be the current (student) model's, where $\theta_i$ denotes the local model parameters being trained and $\tau$ is a temperature hyperparameter (typically $\tau = 2$). The distillation loss is the KL divergence:

\begin{equation}
\mathcal{L}_{\text{distill}} = \tau^2 \cdot \text{KL}(\mathbf{q}_{\text{old}} \| \mathbf{q}_{\text{new}}) = \tau^2 \sum_{c=0}^{C-1} q_{\text{old}}^{(c)} \log\frac{q_{\text{old}}^{(c)}}{q_{\text{new}}^{(c)}}\,
\label{eq:distillation_loss}
\end{equation}
where the $\tau^2$ term compensates for the gradient scaling introduced by the temperature.

\subsection{Multi-Component Loss Function}
\label{sec:loss_function}
The local training objective combines three loss terms: (1) class-reweighted classification loss, (2) knowledge distillation loss, and (3) replay loss on drift-compensated features. The total loss is given by
\begin{equation}
\mathcal{L}_{\text{total}} = \mathcal{L}_{\text{cls}} + \lambda_{\text{distill}}^{(r)} \cdot \mathcal{L}_{\text{distill}} + \lambda_{\text{replay}}^{(r)} \cdot \mathcal{L}_{\text{replay}},
\label{eq:total_loss}
\end{equation}
where $\lambda_{\text{distill}}^{(r)}$ and $\lambda_{\text{replay}}^{(r)}$ are \emph{forgetting-aware adaptive weights} that are dynamically adjusted at each communication round $r$ based on the current forgetting severity (see Section~\ref{sec:adaptive_balancing}).

\subsubsection{Class-Reweighted Classification Loss}

The classification loss is a weighted cross-entropy loss that adjusts the importance of each class based on its frequency and novelty:
\begin{equation}
\mathcal{L}_{\text{cls}} = -\frac{1}{B} \sum_{j=1}^{B} w_{y_j} \cdot \log \frac{\exp(f_{\theta}(x_j)_{y_j})}{\sum_{c=0}^{C-1} \exp(f_{\theta}(x_j)_c)},
\label{eq:classification_loss}
\end{equation}
where $B$ is the batch size, $y_j$ is the label of the $j$-th sample, and $w_{y_j}$ is the class weight from Eq.~\eqref{eq:class_weights}.

\subsubsection{Replay Loss}

Using the drift-compensated embeddings from Eq.~\eqref{eq:feature_compensation}, the replay loss is:
\begin{equation}
\mathcal{L}_{\text{replay}} = -\frac{1}{B_r} \sum_{j=1}^{B_r} \log \frac{\exp(g_{\theta_c}(\tilde{\mathbf{z}}_j)_{y_j})}{\sum_{c=0}^{C-1} \exp(g_{\theta_c}(\tilde{\mathbf{z}}_j)_c)},
\label{eq:replay_loss}
\end{equation}
where $B_r$ is the replay batch size, $\tilde{\mathbf{z}}_j$ is the drift-corrected embedding, and $g_{\theta_c}(\cdot)$ denotes the classifier applied directly to the embedding (bypassing the feature extractor).

\subsection{Class-Aware Federated Aggregation (Level~III)}
\label{sec:federated_mechanism}
After local training, the regional aggregation satellite receives client updates $\{\theta_i^{(r)}, \{\mathbf{p}_c^{(i)}\}, \{n_i^{(c)}\}\}_{i=1}^{N}$ and then aggregates them to form the new global model. Conventional FedAvg uses uniform or data-size-proportional client weights, which under Non-IID data can bias the global classifier toward majority classes: clients with few or no samples of a rare class still contribute their (poorly trained) classifier row for that class, so the global decision boundary for the rare class is diluted by irrelevant clients. We therefore propose \emph{class-aware aggregation}, in which each client's contribution to the \emph{global classifier} is weighted by its \emph{per-class} sample counts. For each class $c$, only clients with $n_i^{(c)} > 0$ participate in aggregating $\mathbf{W}_c^{(G)}$, with weight proportional to $n_i^{(c)}$. The global boundary for $c$ is thus determined solely by clients that actually observed class $c$, so rare classes are no longer diluted and majority-class bias is reduced.

\subsubsection{Feature Extractor Aggregation}

Because the feature extractor $\theta_f$ is shared across all classes, we aggregate it by simple averaging:
\begin{equation}
\theta_f^{(G)} = \frac{1}{N} \sum_{i=1}^{N} \theta_f^{(i)},
\label{eq:feature_aggregation}
\end{equation}
where $\theta_f^{(i)}$ denotes the local feature extractor parameters from client $i$. This ensures that the global feature extractor benefits equally from all clients.

\subsubsection{Class-Aware Classifier Aggregation}

The classifier $\theta_c = \{\mathbf{W}, \mathbf{b}\}$ ($\mathbf{W} \in \mathbb{R}^{C \times d}$, $\mathbf{b} \in \mathbb{R}^C$) is aggregated per class to account for heterogeneous class distributions.

\textbf{Per-Class Aggregation Weights.} For each class $c$, we compute a set of aggregation weights $\{\alpha_i^{(c)}\}_{i=1}^{N}$ that reflect each client's contribution to learning class $c$:
\begin{equation}
\alpha_i^{(c)} = \begin{cases}
\frac{n_i^{(c)}}{\sum_{j=1}^{N} n_j^{(c)}}, & \text{if } n_i^{(c)} > 0, \\
0, & \text{otherwise}.
\end{cases}
\label{eq:class_weights_aggregation}
\end{equation}

If no client has observed class $c$ (i.e., $\sum_{j=1}^{N} n_j^{(c)} = 0$), we fall back to uniform aggregation: $\alpha_i^{(c)} = 1/N$ for all $i$.

\textbf{Classifier Parameter Aggregation.} The global classifier parameters are computed as:
\begin{equation}
\mathbf{W}_c^{(G)} = \sum_{i=1}^{N} \alpha_i^{(c)} \cdot \mathbf{W}_c^{(i)}, \quad b_c^{(G)} = \sum_{i=1}^{N} \alpha_i^{(c)} \cdot b_c^{(i)},
\label{eq:classifier_aggregation}
\end{equation}
for each class $c \in \mathcal{Y}_{1:t}$.

\subsubsection{Communication Efficiency}

To minimize communication overhead, we transmit only model parameters, per-class sample counts, and class prototypes. The communication cost per client per round is
\begin{equation}
\text{CommCost} = |\theta| \cdot 4 \text{ bytes} + |\mathcal{Y}_{1:t}| \cdot (4 + d \cdot 4) \text{ bytes},
\label{eq:comm_cost}
\end{equation}
where $|\theta|$ is the number of model parameters and $d$ is the feature dimension. Since the second term (prototypes and per-class counts) scales only with the number of classes, it is negligible compared to the first term (model parameters), so the additional communication cost introduced by prototype transmission is minimal. Concrete numbers for the model used in our experiments are reported in Section~\ref{sec:resource_overhead}.

\subsubsection{Regional Satellite Aggregation}
\label{alg:server_aggregation}

The aggregation procedure (the ``Class-Aware Aggregation'' block in Algorithm~\ref{alg:federated_cil}) comprises three operations: (1) feature extractor averaging via Eq.~\eqref{eq:feature_aggregation}, (2) class-aware classifier aggregation via Eq.~\eqref{eq:class_weights_aggregation}, and (3) federated prototype aggregation. For the third operation, the server fuses local prototypes into globally consistent class prototypes:
\begin{equation}
\mathbf{p}_c^{(G)} = \frac{1}{N_c} \sum_{i=1}^{N} n_i^{(c)} \cdot \mathbf{p}_c^{(i)},
\label{eq:prototype_aggregation}
\end{equation}
where $N_c = \sum_{i=1}^{N} n_i^{(c)}$ is the total number of class-$c$ samples across all clients. The resulting global prototypes are broadcast to all clients. Algorithm~\ref{alg:federated_cil} summarizes the complete federated training process.

\begin{algorithm}[t]
    \caption{MLFCIL: A Multi-Level Forgetting Mitigation Framework for Federated Class-Incremental Learning in LEO Satellites}
    \label{alg:federated_cil}
    \small
    \begin{algorithmic}[1]
    \REQUIRE Clients $N$, tasks $T$, datasets $\{\mathcal{D}^{(t)}\}$, rounds $E_G$, local epochs $E_L$
    \ENSURE Trained global model $\theta_G^{(T)}$
    \STATE Initialize $\theta_G^{(0)}$, prototypes $\{\mathbf{p}_c^{(G)}\} \leftarrow \emptyset$
    \FOR{task $t = 1$ to $T$}
        \IF{$t > 1$}
            \STATE Save teacher: $\theta_{\text{old}} \leftarrow \theta_G^{(t-1)}$; snapshot prototypes $\hat{\mathbf{p}}_c^{(t)}$
        \ENDIF
        \FOR{round $r = 1$ to $E_G$}
            \STATE Broadcast $\theta_G^{(r-1)}$, $\{\mathbf{p}_c^{(G)}\}$, $\theta_{\text{old}}$ to all clients \hfill $\triangleright$ $\theta_{\text{old}} = \theta_G^{(t-1)}$, fixed throughout task $t$
            \FOR{client $i \in \mathcal{N}$ \textbf{in parallel}}
                \STATE $\theta_i^{(r)}, \{\mathbf{p}_c^{(i)}\}, \{n_i^{(c)}\} \leftarrow$ \textbf{LocalTrain}($\theta_G^{(r-1)}, \{\mathbf{p}_c^{(G)}\}, \hat{\mathbf{p}}^{(t)}, \theta_{\text{old}}, \mathcal{D}_i^{(t)}, t$) \hfill $\triangleright$ Alg.~\ref{alg:client_training}
            \ENDFOR
            \STATE \textbf{// Class-Aware Aggregation}
            \STATE $\theta_f^{(G)} \leftarrow \frac{1}{N}\sum_i \theta_f^{(i)}$ \hfill $\triangleright$ Eq.~\eqref{eq:feature_aggregation}
            \FOR{each class $c \in \mathcal{Y}_{1:t}$}
                \STATE $\alpha_i^{(c)} \leftarrow n_i^{(c)} / \sum_j n_j^{(c)}$; \;$\mathbf{W}_c^{(G)} \leftarrow \sum_i \alpha_i^{(c)} \mathbf{W}_c^{(i)}$
                \STATE $\mathbf{p}_c^{(G)} \leftarrow \sum_i \alpha_i^{(c)} \mathbf{p}_c^{(i)}$
            \ENDFOR
        \ENDFOR
        \STATE Evaluate on $\mathcal{D}_{\text{test}}^{(1:t)}$; record $A_t$
    \ENDFOR
    \end{algorithmic}
\end{algorithm}

\subsection{Dual-Granularity Coordination Strategy}
\label{sec:dual_granularity}

The dual-granularity coordination strategy strengthens the stability-plasticity balance by operating at two complementary time scales: round-level adaptive loss balancing adjusts the relative magnitudes of stability and plasticity signals across communication rounds, while step-level gradient projection resolves their directional conflicts within each training step.

\subsubsection{Forgetting-Aware Adaptive Loss Balancing}
\label{sec:adaptive_balancing}

A critical limitation of existing FCIL methods is the use of fixed loss weights across all incremental tasks, ignoring that forgetting severity varies significantly as the number of learned classes grows. In early tasks with few old classes, excessive distillation constrains plasticity and slows new-class adaptation, whereas in later tasks with many old classes, insufficient distillation leads to severe forgetting. To address this, we propose a \emph{forgetting-aware adaptive loss balancing} mechanism that dynamically adjusts $\lambda_{\text{distill}}$ and $\lambda_{\text{replay}}$ based on a real-time forgetting signal estimated from the replay buffer.

\textbf{Forgetting Signal Estimation.} At the beginning of each global communication round $r$ in task $t > 1$, drift vectors $\boldsymbol{\delta}_c$ (Eq.~\eqref{eq:feature_compensation}) are first computed from the current global prototypes and the task-transition snapshot. Each client $i$ then evaluates the current classifier on the replay buffer under \emph{both} the compensated and the raw feature views, and takes the worse case:
\begin{equation}
\mathcal{F}_i^{(r)} = \max\!\Big(\underbrace{1 - \mathrm{Acc}\!\left(g_{\theta_c^{(i)}},\, \tilde{\mathbf{z}}\right)}_{\text{compensated}},\;\; \underbrace{1 - \mathrm{Acc}\!\left(g_{\theta_c^{(i)}},\, \mathbf{z}\right)}_{\text{raw}}\Big),
\label{eq:forgetting_score}
\end{equation}
where $\mathrm{Acc}(g, \mathbf{v}) = \frac{1}{|\mathcal{M}_i|}\sum_{(\mathbf{v}_j, y_j)} \mathbbm{1}[\arg\max_c\, g(\mathbf{v}_j){=}y_j]$, $\tilde{\mathbf{z}}_j = \mathbf{z}_j + \boldsymbol{\delta}_{y_j}$ is the drift-compensated embedding, and $\mathbf{z}_j$ is the raw stored embedding.
Taking the maximum of the two views serves as a conservative lower bound on retention: the compensated view captures residual forgetting after drift correction (matching the replay loss), while the raw view guards against the scenario where prototype-level drift estimation underestimates per-sample feature shift, ensuring that the adaptive weights never underreact to forgetting.
$\mathcal{F}_i^{(r)} \in [0, 1]$, with values near 0 indicating strong retention and values near 1 indicating severe forgetting.

\textbf{Dynamic Weight Adjustment.} The loss weights are adapted as a function of the forgetting score:
\begin{equation}
\lambda_{\text{distill}}^{(r)} = \lambda_{\text{distill}}^{0} \cdot \left(1 + \gamma \cdot \mathcal{F}_i^{(r)}\right), \quad
\lambda_{\text{replay}}^{(r)} = \lambda_{\text{replay}}^{0} \cdot \left(1 + \gamma \cdot \mathcal{F}_i^{(r)}\right),
\label{eq:adaptive_weights}
\end{equation}
where $\lambda_{\text{distill}}^{0} = 0.5$ and $\lambda_{\text{replay}}^{0} = 0.3$ are base weights, and $\gamma = 2.0$ is a sensitivity parameter controlling the adaptation strength. To prevent excessively large weights from destabilizing training, the adapted weights are clipped:
\begin{equation}
\lambda_{\text{distill}}^{(r)} = \min\left(\lambda_{\text{distill}}^{(r)},\; \lambda_{\text{distill}}^{\max}\right), \quad
\lambda_{\text{replay}}^{(r)} = \min\left(\lambda_{\text{replay}}^{(r)},\; \lambda_{\text{replay}}^{\max}\right),
\label{eq:weight_clipping}
\end{equation}
where $\lambda_{\text{distill}}^{\max} = 1.5$ and $\lambda_{\text{replay}}^{\max} = 1.0$.

Computing $\mathcal{F}_i^{(r)}$ requires only a single forward pass through the classifier on the replay buffer, adding negligible cost. When forgetting is high, the weights $\lambda_{\text{distill}}$ and $\lambda_{\text{replay}}$ increase to strengthen anti-forgetting supervision. When retention is satisfactory, they decrease to favor plasticity. This adaptive behavior is particularly important in satellite FCIL, where forgetting severity varies across tasks due to Non-IID and geographic heterogeneity.

\subsubsection{Gradient Projection for Conflict Resolution}

\begin{figure}[t]
\centering
\includegraphics[width=0.45\textwidth]{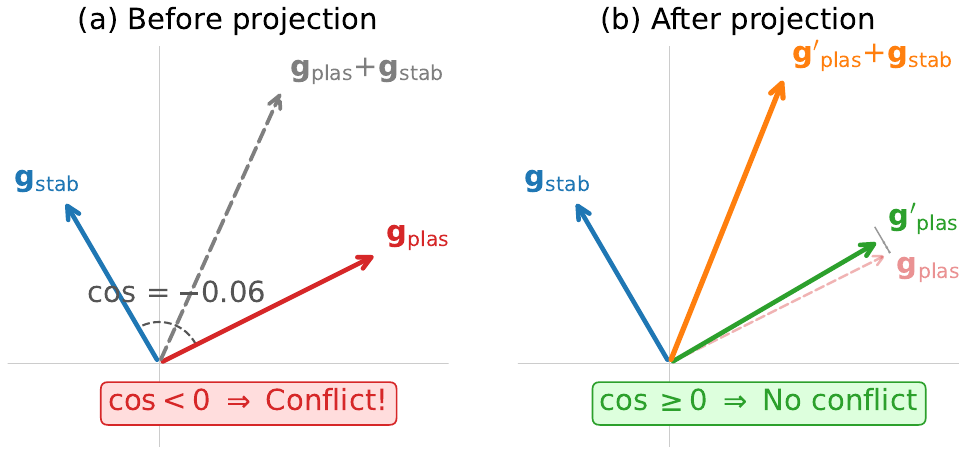}
\caption{Stability-plasticity gradient projection. (a)~Before projection (conflict). (b)~After projection (resolved).}
\label{fig:gradient_projection}
\end{figure}

We decompose the total gradient on $\theta_f$ into plasticity and stability components:
\begin{equation}
\mathbf{g}_{\text{plas}} = \nabla_{\theta_f} \mathcal{L}_{\text{cls}}, \quad
\mathbf{g}_{\text{stab}} = \nabla_{\theta_f} \left(\lambda_{\text{distill}}^{(r)} \mathcal{L}_{\text{distill}}\right).
\label{eq:gradient_decompose}
\end{equation}
Note that $\mathcal{L}_{\text{replay}}$ does not contribute gradients to $\theta_f$, since replay is applied at the classifier input. A conflict is detected when
\begin{equation}
\cos(\mathbf{g}_{\text{plas}}, \mathbf{g}_{\text{stab}}) = \frac{\mathbf{g}_{\text{plas}} \cdot \mathbf{g}_{\text{stab}}}{\|\mathbf{g}_{\text{plas}}\| \|\mathbf{g}_{\text{stab}}\|} < 0.
\label{eq:conflict_detection}
\end{equation}
Such conflicts occur frequently in the early rounds of each incremental task, when the model is adapting rapidly to new data while the distillation loss resists changes to the feature extractor. When a conflict is detected (Fig.~\ref{fig:gradient_projection}(a)), we resolve it by projecting the plasticity gradient so that it no longer opposes the stability gradient (Fig.~\ref{fig:gradient_projection}(b)), following the principle that new-class learning should not harm old-class preservation. Formally,

\begin{equation}
\mathbf{g}_{\text{plas}}^{\prime} = \begin{cases}
\mathbf{g}_{\text{plas}} - \frac{\mathbf{g}_{\text{plas}} \cdot \mathbf{g}_{\text{stab}}}{\|\mathbf{g}_{\text{stab}}\|^2} \cdot \mathbf{g}_{\text{stab}}, & \text{if } \mathbf{g}_{\text{plas}} \cdot \mathbf{g}_{\text{stab}} < 0, \\
\mathbf{g}_{\text{plas}}, & \text{otherwise},
\end{cases}
\label{eq:gradient_projection}
\end{equation}
so that $\mathbf{g}_{\text{plas}}^{\prime}$ is at worst orthogonal to $\mathbf{g}_{\text{stab}}$. The resulting update for $\theta_f$ is
\begin{equation}
\theta_f \leftarrow \theta_f - \eta \left(\mathbf{g}_{\text{plas}}^{\prime} + \mathbf{g}_{\text{stab}}\right),
\label{eq:coordinated_update}
\end{equation}
where $\eta$ is the learning rate. For the classifier parameters $\theta_c$, we apply the standard weighted combination of gradients, since the replay loss provides direct supervision on the classifier.

\subsubsection{Multi-Level Coordination Design Rationale}

The overall coordination design follows three principles:
\begin{itemize}
    \item \textbf{Separation of concerns.} The feature extractor $\theta_f$ is treated as a \emph{shared representation} that needs to balance plasticity and stability, while the classifier $\theta_c$ is treated as a \emph{task-specific head} with stronger old-class protection from both KD and replay.
    \item \textbf{Conflict resolution at the gradient level.} When plasticity and stability gradients directly oppose each other on $\theta_f$, the projection (Eq.~\eqref{eq:gradient_projection}) ensures new-class learning proceeds in directions that are at worst neutral to old-class preservation, rather than harmful.
    \item \textbf{Adaptive feedback at the loss level.} The forgetting-aware balancing (Eq.~\eqref{eq:adaptive_weights}) adjusts the \emph{relative magnitudes} of stability vs.\ plasticity signals across rounds, while the gradient projection handles their \emph{directional conflicts} within each step. These two mechanisms operate at complementary granularities (round-level magnitude vs.\ step-level direction) and jointly realize effective coordination.
\end{itemize}
Table~\ref{tab:coordination} summarizes which parameter group each mechanism updates and where conflicts are resolved.

\begin{table}[t]
\centering
\caption{Which mechanisms update which parameter group ($\theta_f$ vs.\ $\theta_c$), and where gradient conflicts are resolved (projection on $\theta_f$).}
\label{tab:coordination}
\small
\begin{tabular}{lcc}
\hline
\textbf{Mechanism} & $\theta_f$ (\textbf{Extractor}) & $\theta_c$ (\textbf{Classifier}) \\
\hline
$\mathcal{L}_{\text{cls}}$ (plasticity) & $\checkmark$ (projected) & $\checkmark$ \\
$\mathcal{L}_{\text{distill}}$ (stability) & $\checkmark$ & $\checkmark$ \\
$\mathcal{L}_{\text{replay}}$ (stability) & --- & $\checkmark$ \\
Adaptive balancing & \multicolumn{2}{c}{adjusts $\lambda$ globally} \\
Gradient projection & $\checkmark$ & --- \\
Class-aware aggr. & --- & $\checkmark$ \\
\hline
\end{tabular}
\end{table}

\begin{algorithm}[t]
    \caption{Client Local Training (\textbf{LocalTrain})}
    \label{alg:client_training}
    \small
    \begin{algorithmic}[1]
    \REQUIRE Global model $\theta_G$, prototypes $\{\mathbf{p}_c^{(G)}\}$, prototype snapshot $\{\hat{\mathbf{p}}_c^{(t)}\}$, teacher $\theta_{\text{old}}$, data $\mathcal{D}_i^{(t)}$, task index $t$
    \ENSURE Updated $\theta_i$, $\{\mathbf{p}_c^{(i)}\}$, $\{n_i^{(c)}\}$
    \STATE $\theta_i \leftarrow \theta_G$; load memory buffer $\mathcal{M}_i$
    \IF{$t > 1$}
        \STATE Compute drift vectors: $\boldsymbol{\delta}_c \leftarrow \mathbf{p}_c^{(G)} - \hat{\mathbf{p}}_c^{(t)}$ for each old class $c$ \hfill $\triangleright$ Eq.~\eqref{eq:feature_compensation}
        \STATE $\mathcal{F}_i^{(r)} \leftarrow \max\!\big(1{-}\mathrm{Acc}(g_{\theta_c},\tilde{\mathbf{z}}),\; 1{-}\mathrm{Acc}(g_{\theta_c},\mathbf{z})\big)$ \hfill $\triangleright$ Eq.~\eqref{eq:forgetting_score}
        \STATE Adapt $\lambda_{\text{distill}}^{(r)}, \lambda_{\text{replay}}^{(r)}$ based on $\mathcal{F}_i^{(r)}$ \hfill $\triangleright$ Eq.~\eqref{eq:adaptive_weights}
    \ELSE
        \STATE $\lambda_{\text{distill}}^{(r)} \leftarrow 0$; $\lambda_{\text{replay}}^{(r)} \leftarrow 0$
    \ENDIF
    \FOR{epoch $e = 1$ to $E_L$}
        \FOR{mini-batch $\mathcal{B} \subset \mathcal{D}_i^{(t)}$}
            \STATE Extract features $\mathbf{z}_j = h_{\theta_f}(x_j)$; update prototypes $\mathbf{p}_{y_j}^{(i)}$ and buffer $\mathcal{M}_i$ \hfill $\triangleright$ Eq.~\eqref{eq:prototype_update}
            \STATE $\mathcal{L}_{\text{cls}} \leftarrow$ weighted cross-entropy with class weights $w_c$ \hfill $\triangleright$ Eq.~\eqref{eq:classification_loss}
            \STATE $\mathcal{L}_{\text{distill}} \leftarrow 0$; $\mathcal{L}_{\text{replay}} \leftarrow 0$
            \IF{$t > 1$}
                \STATE $\mathcal{L}_{\text{distill}} \leftarrow \tau^2 \cdot \text{KL}(\text{softmax}(f_{\theta_{\text{old}}}(x)/\tau) \| \text{softmax}(f_{\theta_i}(x)/\tau))$ \hfill $\triangleright$ Eq.~\eqref{eq:distillation_loss}
            \ENDIF
            \IF{$|\mathcal{M}_i| > 0$}
                \STATE Sample $\{(\mathbf{z}_k, y_k)\} \sim \mathcal{M}_i$; $\tilde{\mathbf{z}}_k \leftarrow \mathbf{z}_k + \boldsymbol{\delta}_{y_k}$
                \STATE $\mathcal{L}_{\text{replay}} \leftarrow$ cross-entropy on $g_{\theta_c}(\tilde{\mathbf{z}}_k)$ \hfill $\triangleright$ Eq.~\eqref{eq:replay_loss}
            \ENDIF
            \STATE \textbf{// Gradient coordination on $\theta_f$}
            \STATE $\mathbf{g}_{\text{plas}} \leftarrow \nabla_{\theta_f} \mathcal{L}_{\text{cls}}$; \;$\mathbf{g}_{\text{stab}} \leftarrow \nabla_{\theta_f} (\lambda_{\text{distill}}^{(r)} \mathcal{L}_{\text{distill}})$
            \IF{$\mathbf{g}_{\text{plas}} \cdot \mathbf{g}_{\text{stab}} < 0$}
                \STATE $\mathbf{g}_{\text{plas}} \leftarrow \mathbf{g}_{\text{plas}} - \frac{\mathbf{g}_{\text{plas}} \cdot \mathbf{g}_{\text{stab}}}{\|\mathbf{g}_{\text{stab}}\|^2} \mathbf{g}_{\text{stab}}$ \hfill $\triangleright$ Eq.~\eqref{eq:gradient_projection}
            \ENDIF
            \STATE $\theta_f \leftarrow \theta_f - \eta (\mathbf{g}_{\text{plas}} + \mathbf{g}_{\text{stab}})$
            \STATE $\theta_c \leftarrow \theta_c - \eta \nabla_{\theta_c} (\mathcal{L}_{\text{cls}} + \lambda_{\text{distill}}^{(r)} \mathcal{L}_{\text{distill}} + \lambda_{\text{replay}}^{(r)} \mathcal{L}_{\text{replay}})$
        \ENDFOR
    \ENDFOR
    \RETURN $\theta_i$, $\{\mathbf{p}_c^{(i)}\}$, $\{n_i^{(c)}\}$
    \end{algorithmic}
\end{algorithm}

\textbf{Computational Overhead Analysis.} The gradient projection adds only $O(|\theta_f|)$ per step (one dot product and one subtraction), with no additional communication or memory; concrete overhead is reported in Section~\ref{sec:resource_overhead}.

Algorithm~\ref{alg:client_training} summarizes the complete client-side local training procedure, integrating the local training mechanisms (Section~\ref{sec:local_training}), multi-component loss (Section~\ref{sec:loss_function}), and gradient coordination described above.

\section{Experiments}
\label{sec:experiments}

\subsection{Experimental Setup}

\textbf{Dataset.} Experiments are conducted on the NWPU-RESISC45 dataset~\cite{ref36}, a large-scale benchmark for remote sensing scene classification comprising 31,500 images across 45 scene categories (700 images per class) from Google Earth. The categories cover diverse land-use and land-cover types, including residential areas, agricultural lands, transportation infrastructure, industrial facilities, natural landscapes, and specialized structures. This coverage makes NWPU-RESISC45 well suited to evaluating class-incremental learning in satellite edge computing scenarios.

\textbf{Task Configuration.} A class-incremental learning scenario is simulated by partitioning the 45 classes into three sequential tasks of 15 classes each. Task~1 uses 15 classes from fundamental land-cover themes (e.g., residential types, basic infrastructure, agricultural lands). Task~2 adds 15 classes from infrastructure and facility-oriented themes (e.g., industrial, commercial, specialized structures). Task~3 adds 15 classes from specialized terrain and environmental themes (e.g., natural features, specific scenarios). The parenthetical phrases describe the \emph{themes} of the classes in each task, not an exhaustive list of the 15 class names. This partitioning reflects realistic satellite mission evolution. Each task dataset is distributed among $N{=}5$ satellite edge nodes using a Dirichlet distribution with concentration parameter $\alpha_{\mathrm{Dir}}{=}0.5$ (default) to simulate Non-IID data distributions.

\textbf{Implementation Details.} The feature extractor is ResNet-34 pre-trained on ImageNet, with the final fully-connected layer replaced by a 256-dimensional feature mapping layer. Training is performed for $E_G{=}5$ global communication rounds per task and $E_L{=}5$ local epochs per round using the Adam optimizer with learning rate $\eta{=}0.001$. The choice of $E_G{=}5$ per task reflects the limited communication windows typical of satellite–ground links and aligns with resource-constrained federated learning settings where fewer rounds are used, and $E_L{=}5$ is a common setting in the federated learning literature. The memory buffer size is set to $M_{\max}{=}1000$ feature embeddings (${\sim}$1~MB with 32-bit floats). Base loss weights are $\lambda_{\text{distill}}^{0}{=}0.5$ and $\lambda_{\text{replay}}^{0}{=}0.3$, with adaptive sensitivity $\gamma{=}2.0$ and clipping thresholds $\lambda_{\text{distill}}^{\max}{=}1.5$, $\lambda_{\text{replay}}^{\max}{=}1.0$. The distillation temperature and new-class boosting factor are set to $\tau{=}2.0$ and $\beta{=}1.5$, respectively.

\textbf{Experimental Platform.} All experiments were run on a computing cluster equipped with 4 NVIDIA A800 GPUs (80~GB each), simulating modern LEO satellite edge computing platforms. Computational efficiency is evaluated via relative overhead in terms of additional operations, yielding a platform-independent measure of algorithmic complexity.

\textbf{Evaluation Metrics.} Four metrics are adopted: (1) \emph{cumulative accuracy} $A_t$, defined as the classification accuracy on test sets containing all classes learned up to task $t$; (2) \emph{final accuracy} $A_T$, the accuracy after learning all tasks; (3) \emph{performance degradation} $\text{PD}{=}A_1{-}A_T$; and (4) \emph{average incremental accuracy} $\bar{A}{=}\frac{1}{T}\sum_{t=1}^{T}A_t$.

\textbf{Baselines.} To isolate the contribution of each proposed component, the following baselines are compared:
\begin{itemize}
    \item \textbf{FedAvg}~\cite{ref37}: Standard federated averaging without any incremental learning mechanism, serving as the lower bound.
    \item \textbf{FedAvg+KD}: FedAvg augmented with knowledge distillation only, evaluating the isolated effect of distillation-based anti-forgetting.
    \item \textbf{FedAvg+Replay}: FedAvg with feature-level memory replay only, evaluating the isolated effect of replay-based anti-forgetting.
    \item \textbf{Joint Training}: Upper bound (oracle) training on all classes simultaneously, representing the theoretical performance ceiling.
\end{itemize}

\subsection{Overall Performance}
\label{sec:main_results}

\begin{figure}[t]
\centering
\includegraphics[width=0.36\textwidth]{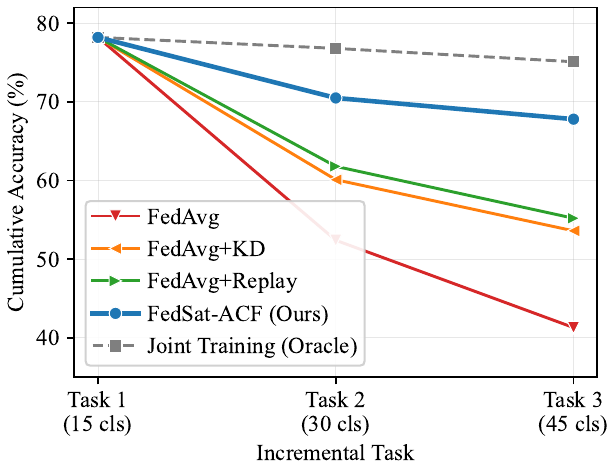}
\caption{Cumulative accuracy across three incremental tasks for all compared methods.}
\label{fig:main_results}
\end{figure}

Table~\ref{tab:main_results} reports the cumulative accuracy across three incremental tasks for all methods under the default Non-IID setting ($\alpha_{\mathrm{Dir}}{=}0.5$), and Fig.~\ref{fig:main_results} illustrates the corresponding performance trajectories.

\begin{table}[t]
\renewcommand{\arraystretch}{1.1}
\centering
\caption{Cumulative accuracy (\%) across incremental tasks on NWPU-RESISC45. $\bar{A}$: average incremental accuracy; PD: performance degradation ($A_1{-}A_T$). Best results (excluding oracle) are in \textbf{bold}.}
\label{tab:main_results}
\footnotesize
\begin{tabular}{lcccccc}
\toprule
\textbf{Method} & $A_1$ & $A_2$ & $A_3$ & $\bar{A}$ & PD$\downarrow$ \\
\midrule
FedAvg                  & 78.2 & 52.4 & 41.3 & 57.3 & 36.9 \\
FedAvg+KD               & 78.2 & 60.1 & 53.6 & 64.0 & 24.6 \\
FedAvg+Replay            & 78.2 & 61.8 & 55.2 & 65.1 & 23.0 \\
\midrule
\textbf{MLFCIL (Ours)} & 78.2 & \textbf{70.5} & \textbf{67.8} & \textbf{72.2} & \textbf{10.4} \\
\midrule
Joint Training (Oracle)  & 78.2 & 76.8 & 75.1 & 76.7 & 3.1 \\
\bottomrule
\end{tabular}
\end{table}

MLFCIL achieves a final accuracy of 67.8\%, significantly outperforming all baselines. Compared to the best single-mechanism baseline, FedAvg+Replay (55.2\%), the proposed method yields a 12.6 percentage point (pp) improvement, indicating that the combined multi-level design is substantially more effective than any individual anti-forgetting strategy. Performance degradation is only 10.4\%, representing a reduction of over 70\% relative to FedAvg (36.9\%). The gap between MLFCIL and the Joint Training oracle (75.1\%) narrows to 7.3~pp, compared to 19.9~pp for FedAvg+Replay, confirming that dual-granularity coordination effectively preserves knowledge across tasks.

\subsection{Ablation Study}
\label{sec:ablation}

\begin{figure}[t]
\centering
\includegraphics[width=0.4\textwidth]{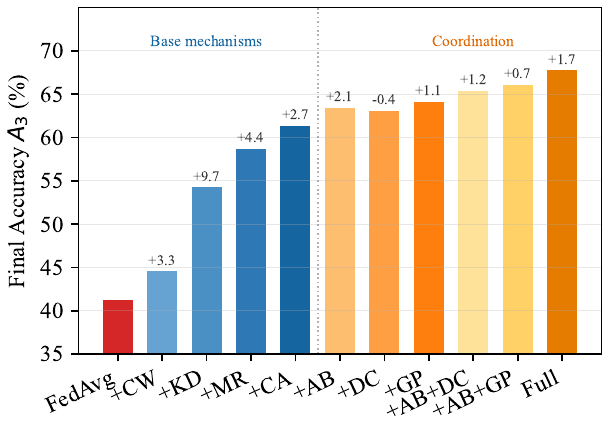}
\caption{Ablation study: final accuracy $A_3$ for each incremental configuration in Table~\ref{tab:ablation}.} 
\label{fig:ablation}
\end{figure}

A comprehensive ablation study is conducted to examine the contribution of each component, including both the base mechanisms and the newly proposed coordination mechanisms. Table~\ref{tab:ablation} and Fig.~\ref{fig:ablation} summarize the results.

\begin{table}[t]
\renewcommand{\arraystretch}{1.1}
\centering
\caption{Ablation study on the NWPU-RESISC45 dataset. CW: class-reweighted loss; KD: knowledge distillation; MR: memory replay; CA: class-aware aggregation; AB: adaptive balancing; DC: drift compensation; GP: gradient projection. ``Full'' denotes MLFCIL with all components.}
\label{tab:ablation}
\footnotesize
\begin{tabular}{lcccc}
\toprule
\textbf{Configuration} & $A_2$ & $A_3$ & $\bar{A}$ & PD$\downarrow$ \\
\midrule
FedAvg (none)           & 52.4 & 41.3 & 57.3 & 36.9 \\
+CW                     & 55.1 & 44.6 & 59.3 & 33.6 \\
+CW+KD                  & 61.8 & 54.3 & 64.8 & 23.9 \\
+CW+KD+MR              & 64.5 & 58.7 & 67.1 & 19.5 \\
+CW+KD+MR+CA           & 66.2 & 61.4 & 68.6 & 16.8 \\
\midrule
+CW+KD+MR+CA+AB        & 67.8 & 63.5 & 69.8 & 14.7 \\
+CW+KD+MR+CA+DC        & 67.5 & 63.1 & 69.6 & 15.1 \\
+CW+KD+MR+CA+GP        & 68.1 & 64.2 & 70.2 & 14.0 \\
+CW+KD+MR+CA+AB+DC     & 69.0 & 65.4 & 70.8 & 12.8 \\
+CW+KD+MR+CA+AB+GP     & 69.4 & 66.1 & 71.2 & 12.1 \\
\midrule
\textbf{Full (all)}     & \textbf{70.5} & \textbf{67.8} & \textbf{72.2} & \textbf{10.4} \\
\bottomrule
\end{tabular}
\end{table}

The ablation study yields the following observations. First, the four base mechanisms (CW, KD, MR, CA) collectively raise $A_3$ from 41.3\% to 61.4\% (+20.1~pp), with KD (+9.7~pp) and MR (+4.4~pp) providing the largest individual gains. Second, adding a single coordination mechanism improves $A_3$ to 63.5\% (AB), 63.1\% (DC), or 64.2\% (GP). Combining two of them (+AB+DC or +AB+GP) yields 65.4\% and 66.1\% respectively, and the full model (all three) reaches 67.8\%, a total gain of 6.4~pp over the base. This gain is notable given the already high baseline, and each mechanism addresses a distinct limitation. Third, the full model outperforms all partial configurations, validating the synergistic design: AB provides adaptive feedback (Task~3 gain: +2.1~pp over the base), DC corrects feature drift for more effective replay (+1.7~pp), and GP resolves gradient conflicts (+2.8~pp).

\subsection{Dual-Granularity Coordination Analysis}
\label{sec:coordination_analysis}

To analyze the proposed coordination mechanisms in depth, three focused analyses are conducted.

\subsubsection{Adaptive Balancing Dynamics}

\begin{figure}[t]
\centering
\includegraphics[width=0.40\textwidth]{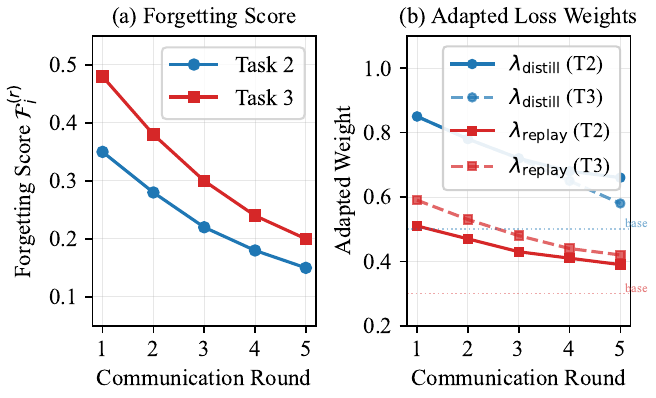}
\caption{Forgetting-aware adaptive balancing dynamics. (a)~Forgetting score $\mathcal{F}_i^{(r)}$ across rounds. (b)~Adapted loss weights $\lambda_{\text{distill}}^{(r)}$ and $\lambda_{\text{replay}}^{(r)}$.}
\label{fig:adaptive_dynamics}
\end{figure}

Fig.~\ref{fig:adaptive_dynamics} shows the evolution of the forgetting score $\mathcal{F}_i^{(r)}$ (Eq.~\eqref{eq:forgetting_score}) and the adapted loss weights across communication rounds for Task~2 and Task~3.

The forgetting score is highest at the beginning of each new task (0.35 for Task~2, 0.48 for Task~3) and decreases as anti-forgetting mechanisms take effect. In early rounds the raw-feature view typically dominates the max because drift compensation has not yet fully converged, providing a conservative protection signal. In later rounds, the two views converge as the feature space stabilizes. The adaptive weights track this signal: $\lambda_{\text{distill}}^{(r)}$ peaks at 0.85 (Task~3, Round~1) before settling to 0.58 by Round~5. Compared to fixed weights ($\lambda_{\text{distill}}{=}0.5$), adaptive balancing provides stronger protection when it is most needed (early rounds) and relaxes constraints in later rounds to improve plasticity.

\subsubsection{Gradient Conflict Frequency and Projection Effect}

\begin{figure}[t]
\centering
\includegraphics[width=0.40\textwidth]{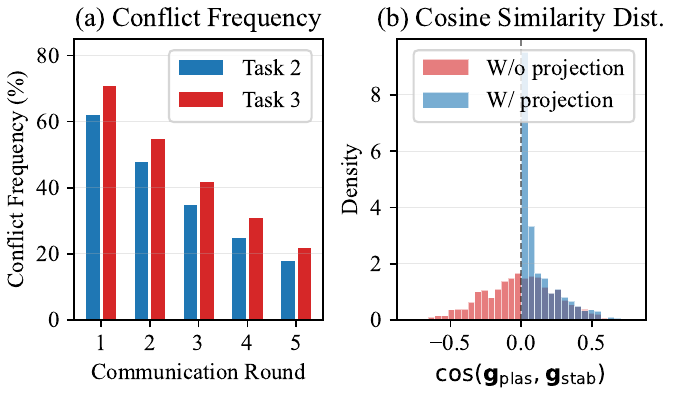}
\caption{Gradient conflict analysis. (a) Frequency of gradient conflicts ($\cos(\mathbf{g}_{\text{plas}}, \mathbf{g}_{\text{stab}})<0$) across training steps. (b) Cosine similarity distribution between plasticity and stability gradients with and without projection.}
\label{fig:gradient_conflict}
\end{figure}

Fig.~\ref{fig:gradient_conflict} analyzes the frequency of gradient conflicts between plasticity and stability on $\theta_f$, and the effect of gradient projection.

Gradient conflicts are prevalent, occurring in 62\% of training steps in Task~2 Round~1 and 71\% in Task~3 Round~1, confirming that na\"ive gradient summation frequently yields suboptimal updates. Without projection, the mean cosine similarity between $\mathbf{g}_{\text{plas}}$ and $\mathbf{g}_{\text{stab}}$ is $-0.12$ (averaged over all steps), indicating a mild but persistent opposition. After projection, the minimum cosine similarity is clamped to zero by construction, and the mean shifts to $+0.08$, so that all updates are at worst neutral with respect to stability.

\subsubsection{Feature Drift Compensation Effect}

\begin{figure}[t]
\centering
\includegraphics[width=0.40\textwidth]{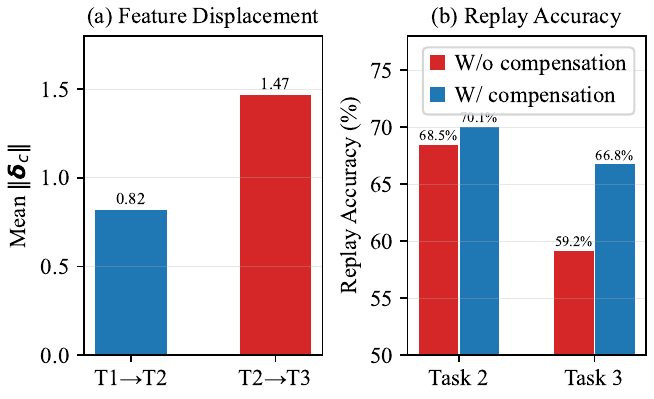}
\caption{Old-class vs.\ new-class accuracy across tasks for MLFCIL and FedAvg.}
\label{fig:drift_compensation}
\end{figure}

Fig.~\ref{fig:drift_compensation} illustrates the effect of prototype-guided drift compensation on replay feature alignment.

The mean feature displacement $\|\boldsymbol{\delta}_c\|$ between stored and current prototypes increases from 0.82 (Task~1$\to$2) to 1.47 (Task~2$\to$3), confirming that feature drift accumulates across task transitions. Without compensation, replay accuracy on old classes drops from 68.5\% (Task~2) to 59.2\% (Task~3) due to misaligned features. With prototype-guided compensation, replay accuracy is maintained at 66.8\% in Task~3, a 7.6~pp improvement. These results validate that the lightweight drift correction ($< 0.01$~ms per sample) effectively maintains replay quality without recomputing stored features.

\subsection{Old vs. New Class Performance}
\label{sec:old_new}

\begin{figure}[t]
\centering
\includegraphics[width=0.40\textwidth]{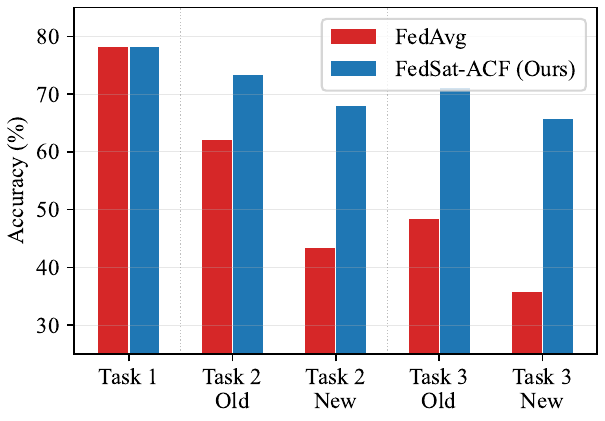}
\caption{Accuracy comparison between old and new classes across tasks. }
\label{fig:old_new}
\end{figure}

Fig.~\ref{fig:old_new} compares accuracy on old classes (learned in previous tasks) with that on new classes (introduced in the current task), to assess the stability and plasticity balance.

MLFCIL maintains old-class accuracy at 73.4\% (Task~2) and 71.2\% (Task~3), with only 2.2~pp degradation across tasks. New-class accuracy reaches 68.1\% (Task~2) and 65.8\% (Task~3), resulting in an old-to-new accuracy gap of 5.3 to 5.4~pp and thus a favorable stability and plasticity balance. In contrast, FedAvg's old-class accuracy drops from 62.1\% to 48.5\% ($-$13.6~pp), indicating severe catastrophic forgetting. The coordination mechanisms contribute to this balance: adaptive balancing adjusts the plasticity and stability trade-off dynamically, while gradient projection prevents new-class learning from directly harming old-class representations.

\subsection{Non-IID Robustness Analysis}
\label{sec:noniid}

A central claim of MLFCIL is robustness under heterogeneous data distributions. Performance is evaluated across Dirichlet concentration parameters $\alpha_{\mathrm{Dir}} \in \{0.1, 0.3, 0.5, 1.0, 5.0\}$, where smaller $\alpha_{\mathrm{Dir}}$ corresponds to stronger Non-IID heterogeneity.

\begin{table}[t]
\renewcommand{\arraystretch}{1.1}
\centering
\caption{Final accuracy $A_3$ (\%) under different Non-IID levels (Dirichlet $\alpha_{\mathrm{Dir}}$). $\Delta$: accuracy drop from $\alpha_{\mathrm{Dir}}{=}5.0$ to $\alpha_{\mathrm{Dir}}{=}0.1$.}
\label{tab:noniid}
\footnotesize
\setlength{\tabcolsep}{4.5pt}
\begin{tabular}{lccccc|c}
\toprule
\textbf{Method} & \multicolumn{5}{c}{Dirichlet $\alpha_{\mathrm{Dir}}$} & \\
\cmidrule(lr){2-6}
 & 0.1 & 0.3 & 0.5 & 1.0 & 5.0 & $\Delta\!\downarrow$ \\
\midrule
FedAvg         & 31.7 & 37.2 & 41.3 & 48.6 & 54.1 & 22.4 \\
FedAvg+KD      & 43.8 & 49.1 & 53.6 & 58.3 & 62.5 & 18.7 \\
FedAvg+Replay  & 45.2 & 50.8 & 55.2 & 59.7 & 63.4 & 18.2 \\
\textbf{Ours}  & \textbf{58.4} & \textbf{63.7} & \textbf{67.8} & \textbf{71.2} & \textbf{73.5} & \textbf{15.1} \\
\bottomrule
\end{tabular}
\end{table}

As shown in Table~\ref{tab:noniid}, MLFCIL consistently outperforms all baselines across all Non-IID levels. The accuracy degradation from near-IID ($\alpha_{\mathrm{Dir}}{=}5.0$) to extreme Non-IID ($\alpha_{\mathrm{Dir}}{=}0.1$) is only 15.1~pp for MLFCIL, compared to 22.4~pp for FedAvg and 18.2 to 18.7~pp for single-mechanism baselines. The class-aware aggregation mechanism is largely responsible for this robustness: under $\alpha_{\mathrm{Dir}}{=}0.1$, where some clients may entirely lack certain classes, per-class weighted aggregation prevents catastrophic bias in the global classifier.

\subsection{Per-Class Performance Under Non-IID}
\label{sec:per_class}

\begin{figure}[t]
\centering
\includegraphics[width=0.40\textwidth]{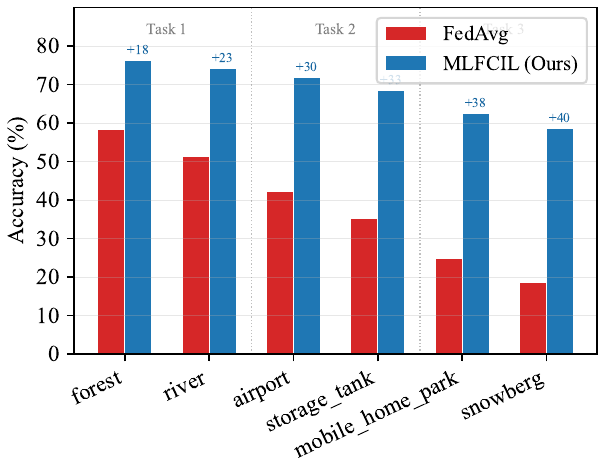}
\caption{Per-class accuracy for six representative classes (two from each of the three tasks), evaluated after all tasks, under Non-IID distributions ($\alpha_{\mathrm{Dir}}{=}0.5$).}
\label{fig:per_class}
\end{figure}

Fig.~\ref{fig:per_class} presents per-class accuracy for six representative classes (two from each of the three tasks), evaluated after all tasks have been learned, under the default Non-IID setting ($\alpha_{\mathrm{Dir}}{=}0.5$).

MLFCIL consistently outperforms FedAvg across all six representative classes, with improvements ranging from 18~pp (\emph{forest}, Task~1) to 40~pp (\emph{snowberg}, Task~3). The gain increases from earlier tasks to later ones, confirming that the multi-level design is particularly effective for classes learned in later incremental stages, where forgetting pressure is highest. Over all 45 classes (not shown), the standard deviation of per-class accuracies is 6.8\% for MLFCIL versus 14.2\% for FedAvg, indicating substantially more uniform performance. Rare classes such as \emph{mobile\_home\_park} and \emph{snowberg} benefit most from class-aware aggregation, which prevents their decision boundaries from being diluted by clients that lack these classes.

\subsection{Memory Efficiency Analysis}
\label{sec:memory_efficiency}

\begin{figure}[t]
\centering
\includegraphics[width=0.40\textwidth]{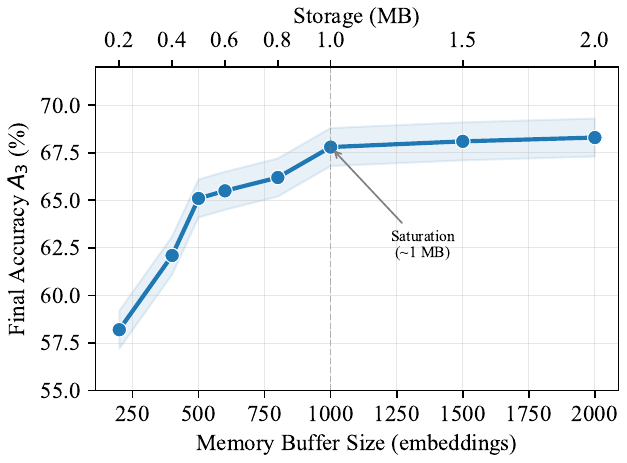}
\caption{Impact of memory buffer size on final accuracy $A_3$.}
\label{fig:memory}
\end{figure}

Fig.~\ref{fig:memory} examines the impact of memory buffer size on performance, with sizes ranging from 200 to 2000 embeddings.

Performance saturates at approximately $M_{\max}{=}1000$ embeddings, corresponding to ${\sim}$1~MB of on-board storage ($256{\times}4$ bytes ${\times}$ 1000). Beyond this threshold, additional memory yields diminishing returns (67.8\% at 1000 vs.\ 68.3\% at 2000). Even with 500 embeddings (0.5~MB), the method achieves 65.1\% accuracy, demonstrating robustness under strict memory constraints.

\begin{table}[t]
    \renewcommand{\arraystretch}{1.1}
    \centering
    \caption{Resource overhead analysis. All overhead percentages are relative to FedAvg.}
    \label{tab:overhead}
    \footnotesize
    \setlength{\tabcolsep}{3pt}
    \begin{tabular}{@{}lll@{}}
    \toprule
    \textbf{Component} & \textbf{Type} & \textbf{Cost} \\
    \midrule
    \multicolumn{3}{@{}l}{\emph{Communication per round}} \\
    \quad Model params & bidir. & ${\sim}$83\,MB \\
    \quad Prototypes (45$\times$256) & up+down & +45\,KB (+0.1\%) \\
    \quad Class counts (45 int.) & upload & +0.2\,KB \\
    \midrule
    \multicolumn{3}{@{}l}{\emph{Computation per step}} \\
    \quad Teacher forward (KD) & 1$\times$ fwd & +2.5\% \\
    \quad Classifier (replay) & 1$\times$ cls-only & +0.5\% \\
    \quad Forgetting score & 1$\times$ cls-only & $<$0.1\% \\
    \quad Gradient projection & dot prod.+sub. & +1.8\% \\
    \quad Drift compensation & vec. addition & $<$0.01\% \\
    \midrule
    \textbf{Total overhead} & & \textbf{+4.9\%} \\
    \midrule
    \multicolumn{3}{@{}l}{\emph{Memory}} \\
    \quad Feature buffer (1000) & on-device & ${\sim}$1\,MB \\
    \quad Teacher model (KD) & on-device & ${\sim}$83\,MB \\
    \bottomrule
    \end{tabular}
\end{table}

\subsection{Resource Overhead Analysis}
\label{sec:resource_overhead}

Table~\ref{tab:overhead} summarizes the communication and computational overhead.

The total computational overhead is only 4.9\% relative to FedAvg, with gradient projection (1.8\%) and KD teacher inference (2.5\%) as the main contributors. Communication overhead is negligible: prototype and statistics transmission adds only 45.2~KB per round ($< 0.12\%$ of total). The memory footprint for the replay buffer (${\sim}$1~MB) and the teacher model (${\sim}$83~MB, one on-device copy of the transmitted model) remains within typical satellite on-board capabilities. These results confirm that MLFCIL is well suited to resource-constrained satellite edge deployment.

\section{Conclusion}
\label{sec:conclusion}

This paper proposed MLFCIL, a multi-level forgetting mitigation framework for federated class-incremental learning in LEO satellites. MLFCIL decomposes catastrophic forgetting into three sources and addresses each at a corresponding level: class-reweighted loss at the class-balanced training level for local training bias, knowledge distillation with feature-level memory replay and prototype-guided drift compensation at the knowledge preservation level for inter-task knowledge loss, and class-aware federated aggregation at the global level for aggregation-induced amplification. A dual-granularity coordination strategy further strengthens the stability-plasticity balance by combining forgetting-aware adaptive loss balancing at each communication round with gradient projection at each training step. Experiments on the NWPU-RESISC45 benchmark under Non-IID satellite scenarios demonstrate that MLFCIL substantially outperforms federated baselines in both cumulative and final accuracy, while incurring minimal computational overhead and requiring only a compact on-device replay buffer. Future work will extend the framework to task-boundary-free online settings and validate it on real satellite constellations with multi-modal observations.


\begin{IEEEbiography}[{\includegraphics[width=1in,height=1.25in,clip,keepaspectratio]{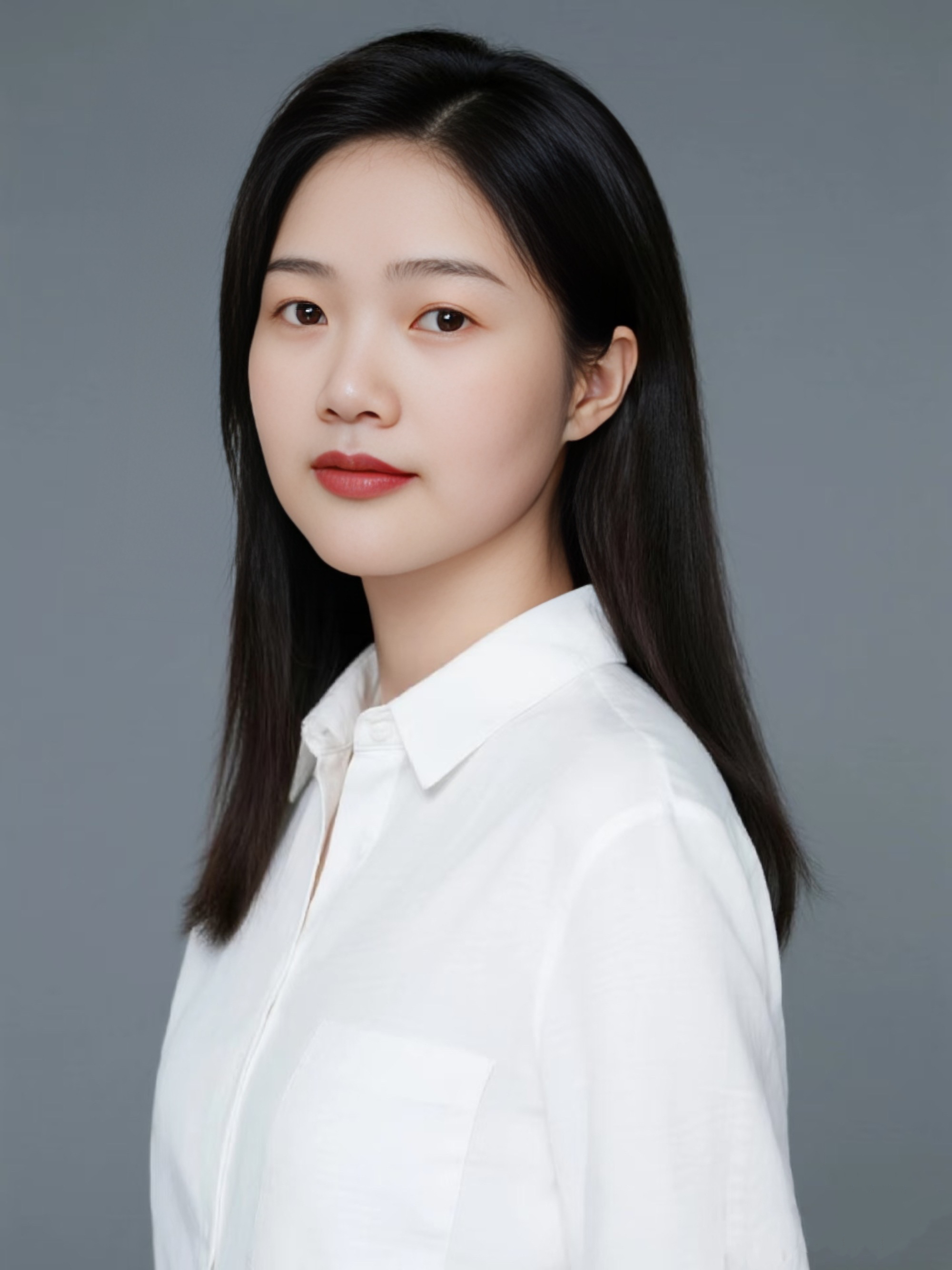}}]{Heng Zhang}
received the M.S. degree in software engineering from Central South University, Changsha, Hunan, China, in 2020. She is currently a teaching assistant with the School of Information Engineering, Gannan University of Science and Technology, Ganzhou, China, as well as a researcher at the Ganzhou Key Laboratory of Cloud Computing and Big Data Research. Her research interests include mobile edge computing, satellite computing, and wireless communications and networking.
\end{IEEEbiography}

\begin{IEEEbiography}[{\includegraphics[width=1in,height=1.25in,clip,keepaspectratio]{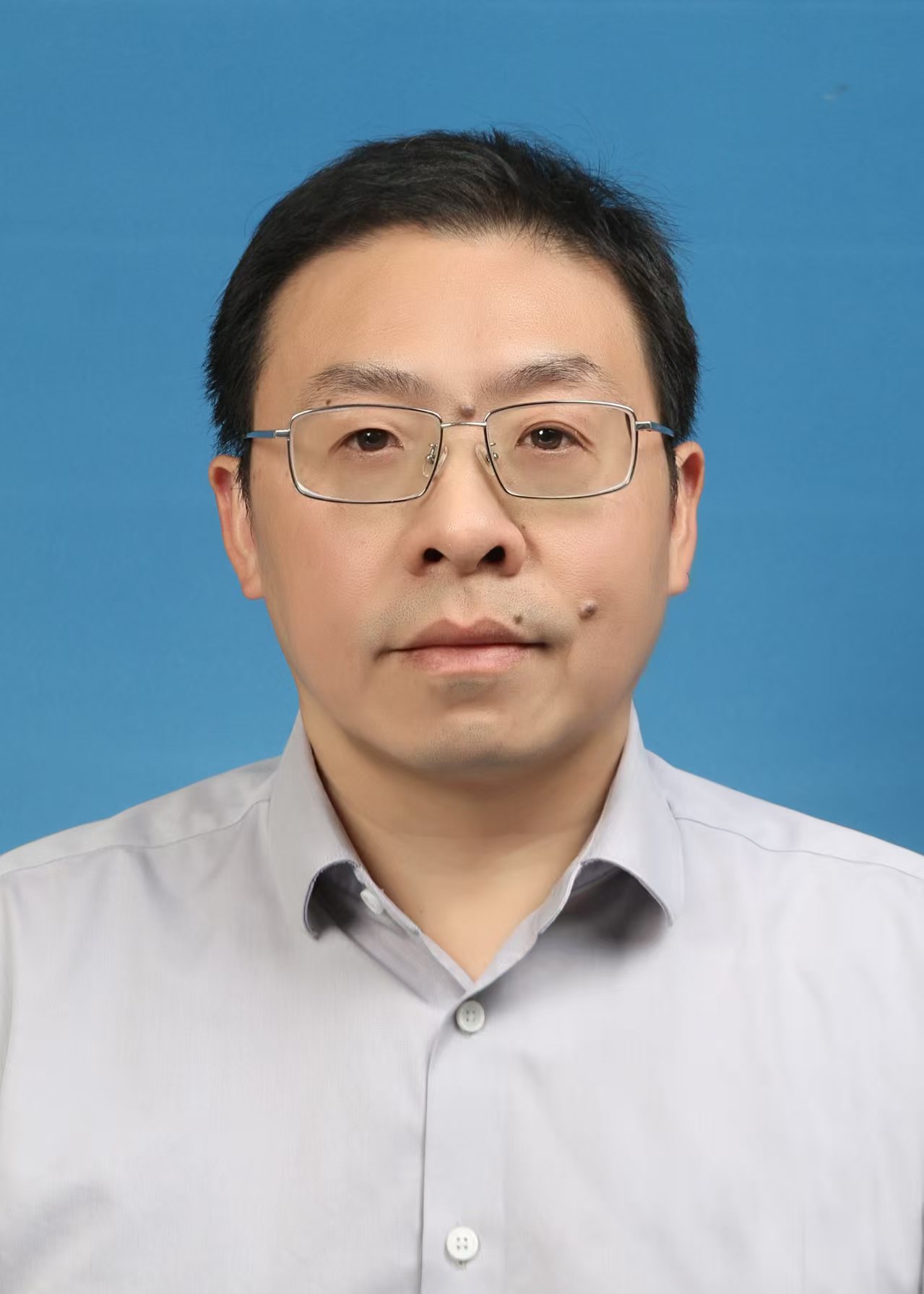}}]{Xiaohong Deng} received the Ph.D. degree in computer science and technology from Central South University, Changsha, China, in 2013. He is currently a Professor with the School of Information Engineering, Gannan University of Science and Technology, and a Master's Supervisor with the Jiangxi University of Science and Technology, Ganzhou. He is also a researcher at the Ganzhou Key Laboratory of Cloud Computing and Big Data Research. His research interests include blockchain, mobile edge computing ,algorithm optimization, and wireless communications and networking. He is a member of China Computer Federation (CCF).
\end{IEEEbiography}

\begin{IEEEbiography}[{\includegraphics[width=1in,height=1.25in,clip,keepaspectratio]{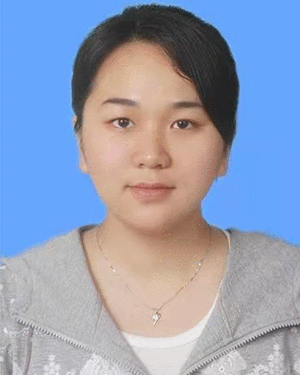}}]{Sijing Duan} (Member, IEEE) received the PhD degree from the School of Computer Science and Engineering, Central South University, Changsha, China, in 2023. She is currently a postdoctoral fellow with the Department of Computer Science and Technology, Tsinghua University, Beijing, China. Her research interests include Big Data analytics, mobile computing, and human mobility modeling.
\end{IEEEbiography}

\begin{IEEEbiography}[{\includegraphics[width=1in,height=1.25in,clip,keepaspectratio]{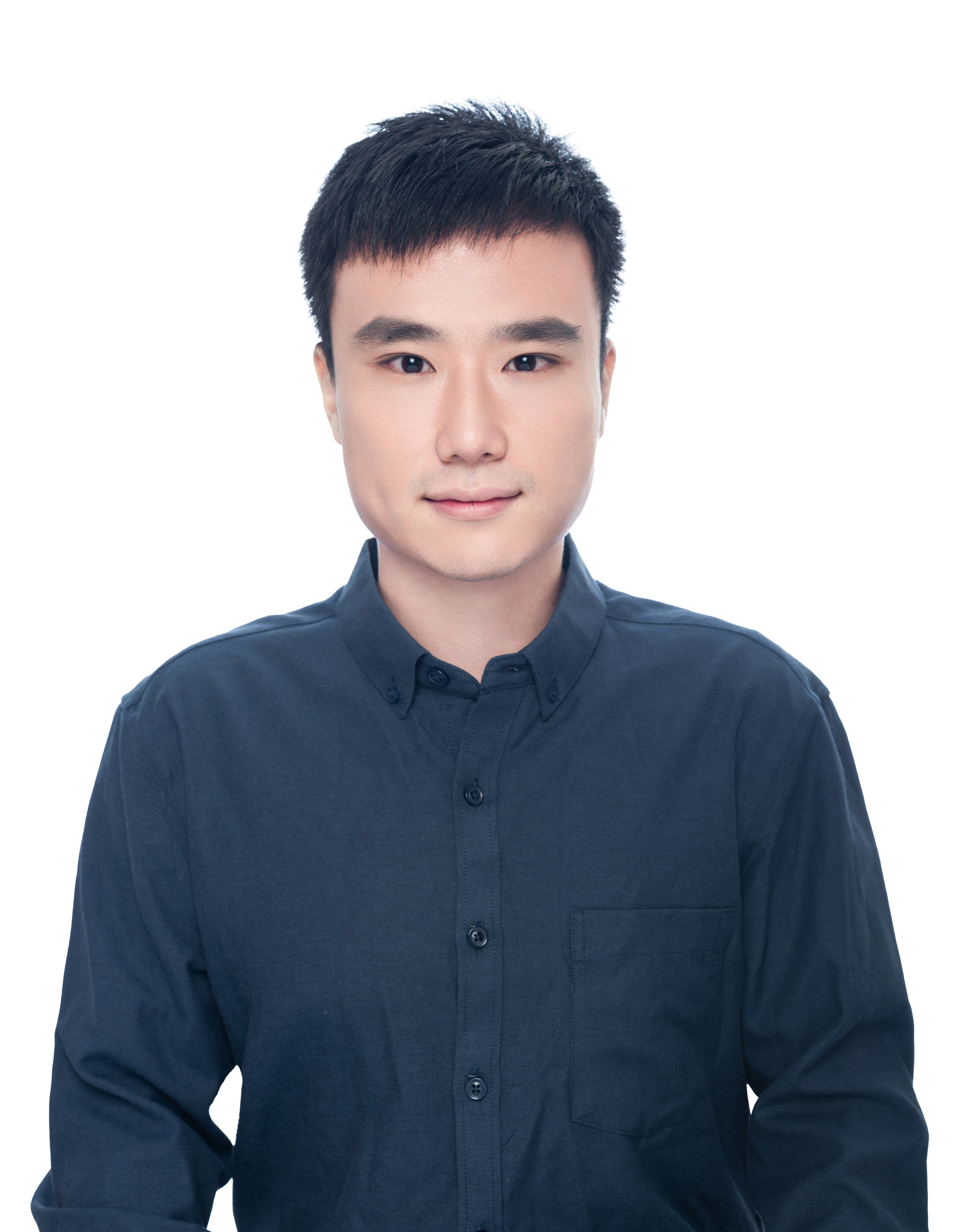}}]{Wu Ouyang} received the M.S. degree in software engineering from Central South University, Changsha, Hunan, China, in 2021. He is currently a teaching assistant with the School of Information Engineering, Gannan University of Science and Technology, Ganzhou, China, as well as a researcher at the Ganzhou Key Laboratory of Cloud Computing and Big Data Research. His research interests include mobile edge computing, satellite computing, and wireless communications and networking.
\end{IEEEbiography}

\begin{IEEEbiography}[{\includegraphics[width=1in,height=1.25in,clip,keepaspectratio]{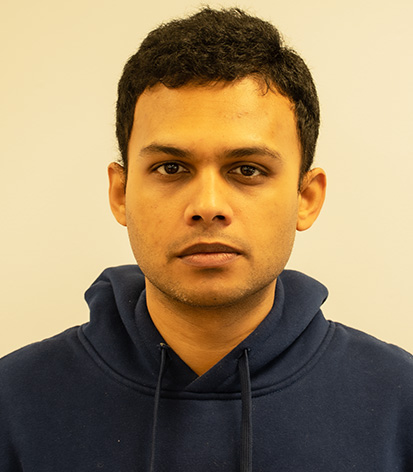}}]{KM Mahfujul} (Member, IEEE) is currently a Post Doctoral Fellow  with the Queen’s University, ON, CA. He received his Ph.D. degree in Electrical and Computer Engineering at Queen's University, ON, Canada, in 2026. Mahfujul received his M.Sc. degree in computer science and technology from Central South University, Hunan, China, in 2020 and his B.Sc. degree in computer science and engineering from Khulna University, Khulna, Bangladesh in 2016. His current research interests include real-time scheduling and learning algorithm design for converged wireless/cloud communication networks.
\end{IEEEbiography}

\begin{IEEEbiography}[{\includegraphics[width=1in,height=1.25in,clip,keepaspectratio]{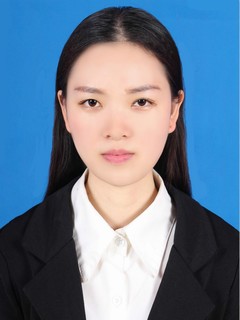}}]{Yiqin Deng} (Member, IEEE) received the M.S. degree in software engineering and the Ph.D. degree in computer science and technology from Central South University, Changsha, China, in 2017 and 2022, respectively. She is currently a Research Assistant Professor with the School of Data Science, Lingnan University, Hong Kong. From 2024 to 2026, she was a Postdoctoral Research Fellow in the Department of Computer Science, City University of Hong Kong. Prior to that, she was a Postdoctoral Research Fellow with the School of Control Science and Engineering, Shandong University, Jinan, China, from 2022 to 2024. She also served as a Visiting Researcher at the University of Florida, Gainesville, FL, USA, from 2019 to 2021. Her research interests include edge computing/AI, wireless communication and networking, computing power networks, and the low-altitude economy.
\end{IEEEbiography}

\begin{IEEEbiography}[{\includegraphics[width=1in,height=1.25in,clip,keepaspectratio]{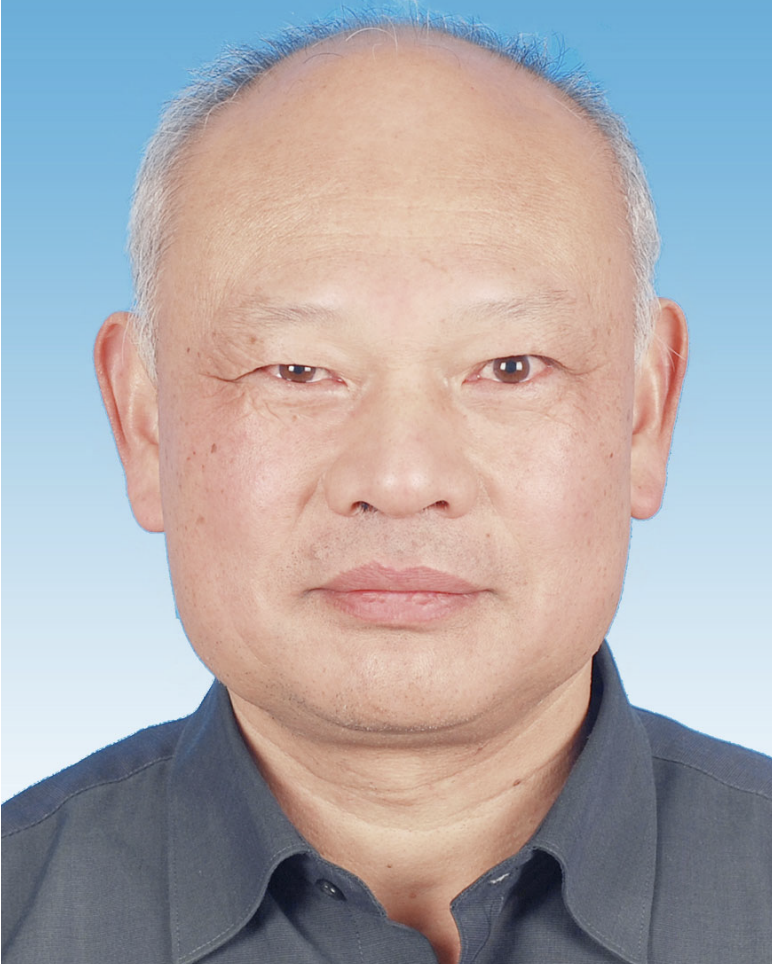}}]{Zhigang Chen} (Member, IEEE) received the B.E., M.S., and Ph.D. degrees from Central South University, China, in 1984, 1987, and 1998, respectively. He is currently a Professor and a Ph.D. Supervisor. His research interests include cluster computing, parallel and distributed systems, computer security, and wireless networks. He is the Director and a Senior Member of China Computer Federation (CCF). He is a member of the Pervasive Computing Committee of CCF.
\end{IEEEbiography}

\vspace{11pt}

\vfill

\end{document}